\documentclass[10pt,journal,compsoc]{IEEEtran}
\ifCLASSOPTIONcompsoc
  \usepackage[nocompress]{cite}
\else
  \usepackage{cite}
\fi
\ifCLASSINFOpdf
\else
\fi

\usepackage{multirow}
\usepackage{booktabs,enumitem,url}
\usepackage{amssymb,amsmath,amsthm}
\usepackage{makecell}
\usepackage{graphicx}
\usepackage{subfigure}
\usepackage{pifont}

\usepackage{amsmath,amsfonts,bm}

\def\eqref#1{equation~\ref{#1}}

\def\1{\bm{1}}

\def\vu{{\bm{u}}}
\def\vv{{\bm{v}}}

\def\vx{{\bm{x}}}

\def\vz{{\bm{z}}}

\DeclareMathAlphabet{\mathsfit}{\encodingdefault}{\sfdefault}{m}{sl}
\SetMathAlphabet{\mathsfit}{bold}{\encodingdefault}{\sfdefault}{bx}{n}

\usepackage{hyperref}
\usepackage{cleveref}
\usepackage{adjustbox}
\usepackage{diagbox}
\usepackage{longtable}
\usepackage[normalem]{ulem}

\usepackage{xcolor}

\begin{document}
%
\title{Geometric Deep Learning for Structure-Based Drug Design: A Survey
%
}
%
%
%
%

\author{Zaixi Zhang*, Jiaxian Yan*, Yining Huang*,
        Qi Liu, Enhong Chen, Mengdi Wang, and Marinka Zitnik
\IEEEcompsocitemizethanks{
\IEEEcompsocthanksitem Zaixi Zhang and Mengdi Wang are with Princeton University. E-mail: $\{$zz8680, mengdiw$\}$@princeton.edu 
\IEEEcompsocthanksitem Jiaxian Yan, Qi Liu, and Enhong Chen are with the Anhui Province Key Laboratory of Big Data Analysis and Application (BDAA), School of Computer Science and Technology, University of Science and Technology of China, Hefei, Anhui 230027, China,  and State Key Laboratory of Cognitive Intelligence, Hefei, Anhui, 230088, China. E-mail: jiaxianyan@mail.ustc.edu.cn, $\{$qiliuql, cheneh$\}$@ustc.edu.cn
\IEEEcompsocthanksitem Yining Huang is from Harvard University. E-mail: yininghuang@hms.harvard.edu
\IEEEcompsocthanksitem Marinka Zitnik is with the Department of Biomedical Informatics, Harvard Medical School; Kempner Institute for the Study of Natural and Artificial Intelligence, Harvard University; Harvard Data Science Initiative; and Broad Institute of MIT and Harvard. E-mail: marinka@hms.harvard.edu.
\IEEEcompsocthanksitem Zaixi Zhang, Jiaxian Yan, and Yining Huang contributed equally to this work. 
\IEEEcompsocthanksitem Correspondence to Qi Liu, Marinka Zitnik, and Mengdi Wang.
}
\thanks{Manuscript received xx xxx. 201x; accepted xx xxx. 201x. Date of Publication xx xxx. 201x; date of current version xx xxx. 201x. This research was partially supported by a grant from the
National Natural Science Foundation of China (Grant No. 61922073). }
}

%
%

\markboth{Journal of \LaTeX\ Class Files,~Vol.~14, No.~8, August~2015}%
{Shell \MakeLowercase{\textit{et al.}}: Bare Demo of IEEEtran.cls for Computer Society Journals}
%



\IEEEtitleabstractindextext{%
\begin{abstract}
Structure-based drug design (SBDD) leverages the three-dimensional geometry of proteins to identify potential drug candidates. Traditional approaches, rooted in physicochemical modeling and domain expertise, are often resource-intensive. Recent advancements in geometric deep learning, which effectively integrate and process 3D geometric data, alongside breakthroughs in accurate protein structure predictions from tools like AlphaFold, have significantly propelled the field forward. This paper systematically reviews the state-of-the-art in geometric deep learning for SBDD. We begin by outlining foundational tasks in SBDD, discussing prevalent 3D protein representations, and highlighting representative predictive and generative models. Next, we provide an in-depth review of key tasks, including binding site prediction, binding pose generation, de novo molecule generation, linker design, protein pocket generation, and binding affinity prediction. For each task, we present formal problem definitions, key methods, datasets, evaluation metrics, and performance benchmarks. Lastly, we explore current challenges and future opportunities in SBDD. Challenges include oversimplified problem formulations, limited out-of-distribution generalization, biosecurity concerns related to the misuse of structural data, insufficient evaluation metrics and large-scale benchmarks, and the need for experimental validation and enhanced model interpretability. Opportunities lie in leveraging multimodal datasets, integrating domain knowledge, developing comprehensive benchmarks, establishing criteria aligned with clinical outcomes, and designing foundation models to expand the scope of design tasks. We also curate \url{https://github.com/zaixizhang/Awesome-SBDD}, reflecting ongoing contributions and new datasets in SBDD.
\end{abstract}
\begin{IEEEkeywords}
Geometric Deep Learning, Generative Models, Molecular Design, Structure-based Drug Design, Therapeutic Science 
\end{IEEEkeywords}}

\maketitle

\IEEEdisplaynontitleabstractindextext

\IEEEpeerreviewmaketitle

\section{Introduction} 
\label{sec:intro}

Structure-based drug design (SBDD) \cite{schneuing2022structure, luo20213d, isert2023structure} aims at designing small molecules that bind to target proteins to inhibit certain protein functions. SBDD leverages the three-dimensional geometric information of target proteins to design and optimize small-molecule drug candidates. 

Traditionally, 3D structures of the target protein are obtained in vitro with techniques like X-ray crystallography \cite{drenth2007principles}, nuclear magnetic resonance (NMR) spectroscopy \cite{mitchell2008nuclear}, or cryo-electron microscopy (cryo-EM) \cite{danev2019cryo}. Recently, the progress on high-accurate protein structure prediction such as AlphaFold \cite{jumper2021highly} and ESMFold \cite{lin2023evolutionary} allows efficient generation of target protein structures in silico, which further boosts the availability of structural data and lays the foundation for SBDD's broad applications.

SBDD has emerged as an instrumental approach in the development of new kinds of therapies. Several drugs available in the market today owe their genesis to SBDD. For instance, HIV-1 protease inhibitors were identified using this approach \cite{wlodawer1998inhibitors}. Similarly, raltitrexed, a thymidylate synthase inhibitor \cite{anderson2003process}, and the antibiotic norfloxacin \cite{rutenber1996binding} are other exemplars of successful SBDD applications. Nevertheless, conventional SBDD methodologies, which are grounded in physical modeling, the application of hand-engineered scoring functions, and exhaustive search across vast biochemical space, present substantial challenges. In response to these limitations, there has been growing interest in geometric deep learning \cite{atz2021geometric} to accelerate and enhance structure-based drug design (SBDD). In this survey, we aim to summarize recent advances in geometric deep learning for SBDD. Given the page constraints, \textbf{we focus our scope on tasks related to the protein-ligand binding interface}, including \emph{ binding site prediction, binding pose generation, de novo ligand generation, linker design, protein pocket generation, and binding affinity prediction} (Figure \ref{sbdd tasks}).

Geometric deep learning (GDL) \cite{atz2021geometric,li2022graph} refers to neural network architectures, including CNN, GNN, and Transformer, designed to capture and encode 3D geometric data. Unlike traditional methods that rely on hand-crafted feature engineering, GDL can autonomously extract salient 3D structural features. Additionally, certain GDL techniques, such as TFN \cite{thomas2018tensor}, EGNN \cite{satorras2021n}, and GMN \cite{huang2022equivariant} that 
integrate symmetry properties directly into the network design, which serves as an effective inductive bias and potentially leads to enhanced performance. In the realm of 3D Euclidean space, "symmetry" encompasses transformations like rotations, translations, and reflections. It's imperative to understand how protein and molecular properties vary under these transformations (Section~\ref{symmetry}). 
With the rapid development of geometric deep learning, a series of SBDD tasks including binding site prediction \cite{gainza2020deciphering}, binding pose generation \cite{corso2022diffdock}, \emph{de novo} ligand generation \cite{peng2022pocket2mol}, linker design \cite{huang20223dlinker}, binding affinity prediction \cite{li2021structure}, and more \cite{dauparas2022robust, watson2023novo} have benefited. 
Geometric deep learning for SBDD has advanced rapidly, drawing more attention from broad communities. Therefore, writing a survey summarizing the recent progress and envisioning the future directions is necessary.

Geometric deep learning for SBDD has two main categories of tasks: predictive \cite{10.1145/3447548.3467311} and generative tasks \cite{luo20213d}. Predictive tasks are concerned with predicting outcomes based on given input small molecule and target protein data (e.g., binding affinity prediction), demanding high accuracy and reliability due to their applications in drug discovery. On the other hand, generative tasks are centered around the design of new small molecule data, such as \emph{de novo} ligand generation. In this paper, we discuss both tasks with an emphasis on exploring the potential and capabilities of the new emerging generative models for SBDD.

\subsection{Distinctive Contributions of the Survey}

Artificial intelligence is increasingly used to augment all stages of scientific research~\cite{wang2023scientific, zhang2023artificial}, including designing and developing new therapeutics. We here focus on SBDD, a critical element of therapeutic science underscored by a plethora of surveys detailing its advancements \cite{verlinde1994structure, colman1994structure, blundell1996structure, klebe2000recent, anderson2003process, ferreira2015molecular, ozccelik2022structure}. However, a common thread among these surveys is their vantage point rooted firmly in biochemistry, which may only partially cater to the machine learning research community. Marking a departure from these extant surveys, our review endeavors to bridge this gap. We overview geometric deep-learning methods explicitly tailored for SBDD, elucidated through the lens of machine learning and deep learning paradigms. A standout feature of our review is its meticulous organization, deeply anchored in SBDD task-specific frameworks. More explicitly, we have curated our sections based on distinct SBDD task categories. We have framed each task within a machine learning challenge, ensuring a seamless marriage between domain-specific intricacies and computational methodologies. This entails clearly presenting algorithms, benchmark datasets, evaluation metrics, and model performances for each task. Through this approach, our aim is two-fold: firstly, to enable researchers from the machine learning and deep learning fraternities to gain insights into SBDD tasks without being burdened by intricate domain-specific prerequisites, and secondly, to lay the groundwork that could motivate the development of more sophisticated geometric deep learning algorithms optimally suited for structure-based drug design.



\subsection{Organization of the Survey}

In this survey, we delve into the interdisciplinary domain bridging geometric deep learning and SBDD. To ensure comprehensive coverage, we curated papers from machine learning and scientific conferences and journals, including NeurIPS, ICLR, ICML, KDD, TKDE, and TPAMI, Nature Machine Intelligence, Nature Computational Science, and Nature Communications. Concurrently, we retrieved publications from natural science journals. Guided by SBDD tasks highlighted in this survey, our search strategy was anchored by key terms, including "structure-based drug design," "protein-ligand docking," "protein-ligand affinity prediction," "linker design," and "protein binding site prediction," among others, ensuring we amassed a broad spectrum of relevant studies. Given that the majority of FDA-approved drugs fall into the category of small molecules \cite{benedetto2022fda}, our focus in this survey primarily centers on papers and methodologies that discuss these small molecules. Following our exhaustive search, we meticulously categorized the gathered papers, aligning them with their respective tasks and methodologies. Table~\ref{tab:sbdd} provides a detailed breakdown.


In Section~\ref{sec:preliminary}, we introduce 3D representations of target proteins, the SBDD tasks reviewed in this paper, and popular predictive and generative models.
The following sections are organized based on the logical order of SBDD tasks as shown in Figure~\ref{sbdd tasks}: as for the target protein structure, we first need to identify the binding site; then we can conduct binding pose generation, \emph{de novo} molecule generation, and linker design; meanwhile, we can optimize the pocket with protein pocket generation; with the protein-ligand complex structure, we can use binding affinity prediction and other filtering criteria to filter drug candidates.
Admittedly, the order and the category of SBDD tasks are not fixed since SBDD is an iterative process that proceeds through multiple cycles, leading optimized drug candidates to clinical trials \cite{batool2019structure}. Some methods may also be capable of accomplishing multiple tasks. For example, EquiBind \cite{stark2022equibind} can predict the binding pose of ligands without prior knowledge of the binding site, i.e., blinding docking, to ease readers' understanding.
Finally, Section~\ref{sec:future} identifies the open challenges and opportunities, paving the way for the future of designing geometric deep learning methods for structure-based drug design.


\section{Preliminaries}

\begin{figure*}[htbp]
\centering
\includegraphics[width=.9\linewidth]{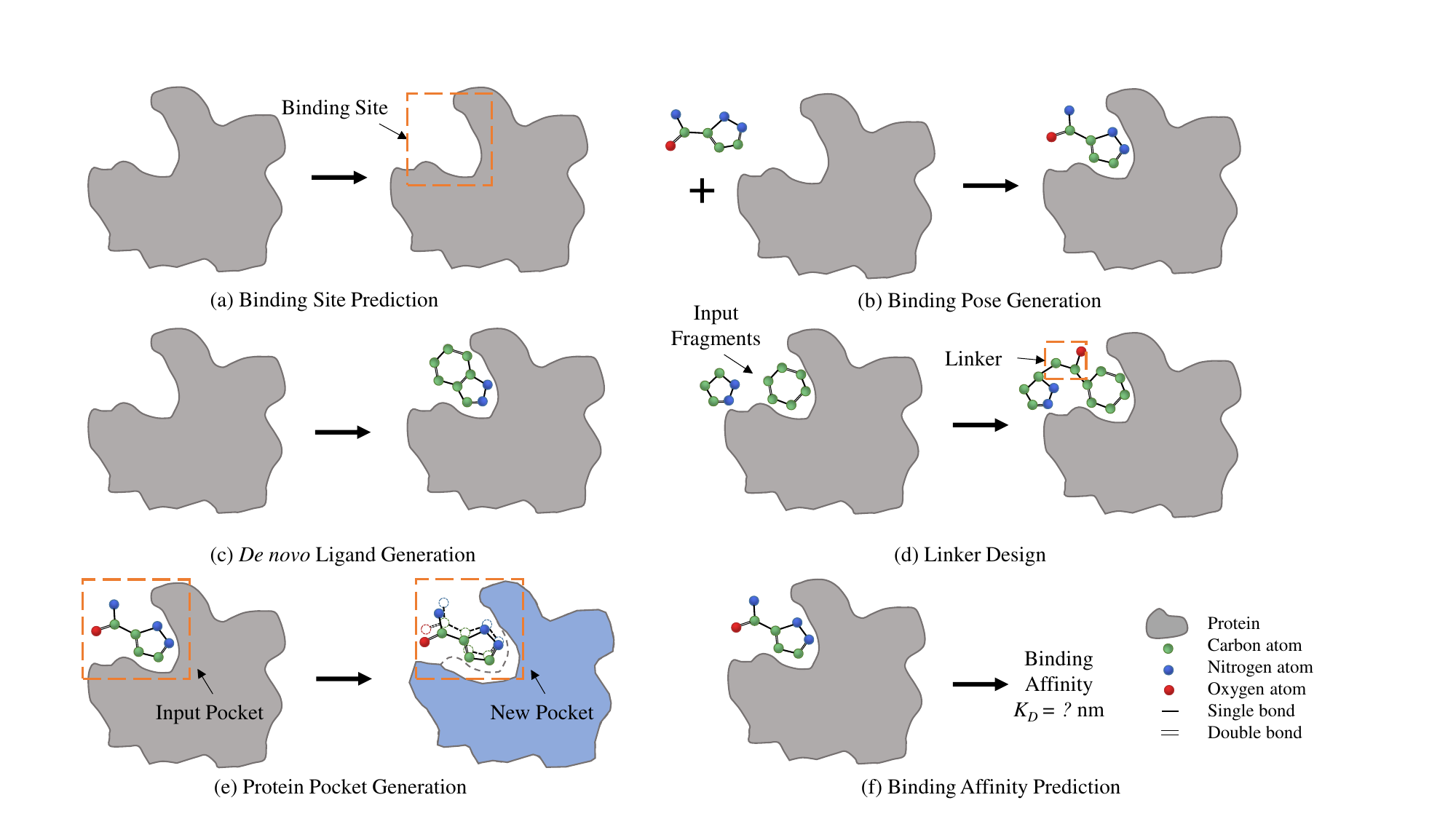}
\caption{Structure-based drug design tasks discussed in this survey: (a) binding site prediction identifies areas of the protein structure that can act as
binding sites for ligands (Section~\ref{sec:binding_site}); (b) binding pose generation or protein-ligand docking focus on predicting the binding conformations of the protein-ligand complex (Section~\ref{sec:binding_pose});
(c) \emph{de novo} ligand generation designs binding
ligands from scratch with the structural information of the target protein (Section~\ref{sec:ligand_generation}); (d) linker design combines disconnected molecular fragments into a combined ligand molecule conditioned on the target protein (Section~\ref{sec:linker_design}); (e) protein pocket generation redesigns the protein pocket (including sequence and structure) given binding ligand (Section~\ref{sec:pocket_generation}); (f) binding affinity prediction predicts the affinity between a protein and a ligand given their binding structure (Section~\ref{sec:binding_affinity}).}
\label{sbdd tasks}
\end{figure*}

Figure~\ref{sbdd tasks} provides an overview of the SBDD tasks encompassing binding site prediction, binding pose generation, \emph{de novo} molecule generation, linker design, and binding affinity prediction. Binding site and affinity prediction are typically formulated as predictive tasks, whereas binding pose generation, \emph{de novo} molecule generation, and linker design are addressed as generative tasks. 

\subsection{Protein and Ligand Representations} 
\label{sec:preliminary}
\begin{figure}[htbp]
\centering
\begin{subfigure}[3D grid]
{\includegraphics[width=.30\linewidth]{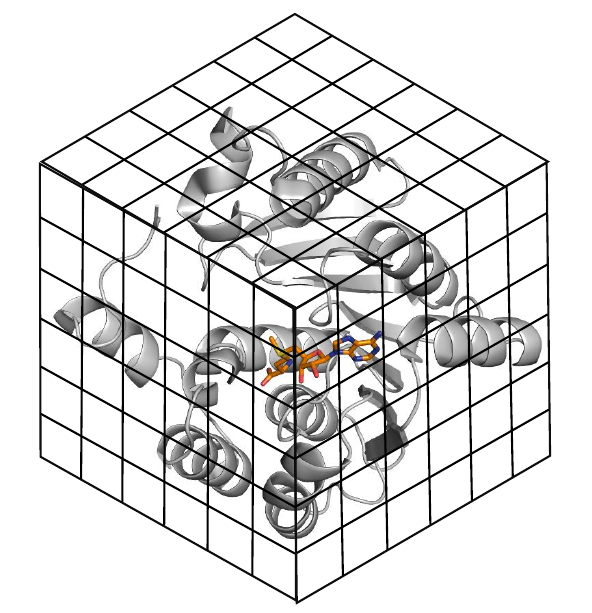} }
\end{subfigure}
\begin{subfigure}[3D surface]
{\includegraphics[width=.30\linewidth]{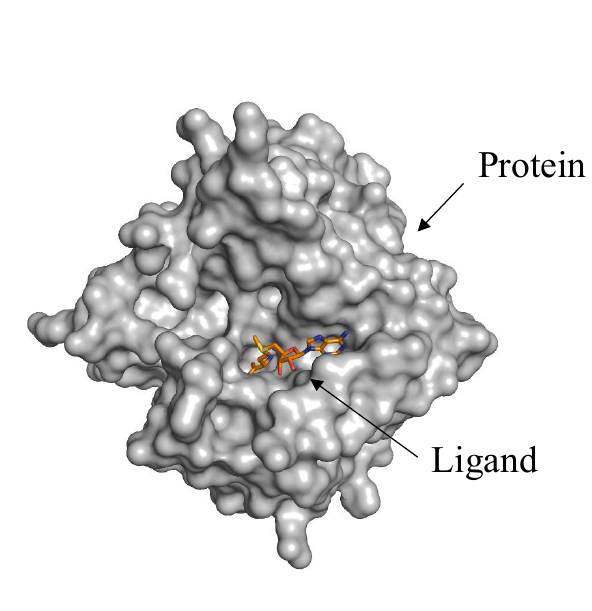} }
\end{subfigure}
\begin{subfigure}[3D graph]
{\includegraphics[width=.30\linewidth]{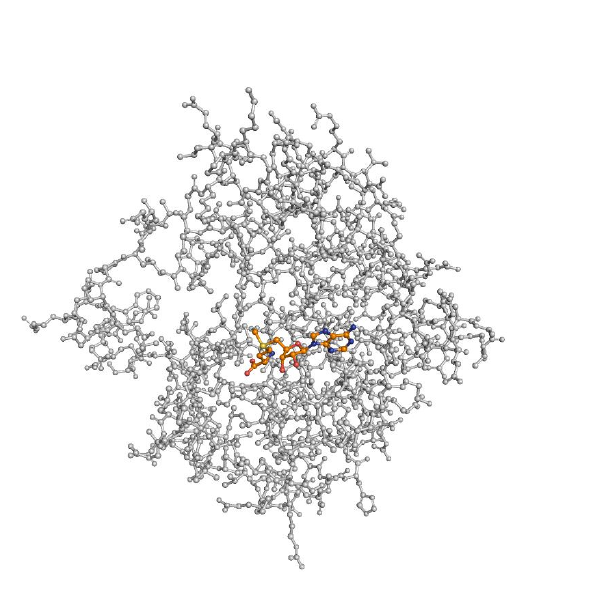} }
\end{subfigure}
\caption{3D representations of proteins used for geometric deep learning: (a) 3D grid, (b) 3D surface, and (c) 3D graph, illustrated for PDB ID 2avd.}
\label{protein representation}
\end{figure}

Protein and ligand molecule representations depend on the SBDD tasks and specific geometric deep learning methods. In SBDD, proteins are usually characterized by 3D representations that capture critical 3D structural information in the form of grids, surfaces, and spatial graphs (Figure~\ref{protein representation}):
\begin{itemize}
\item 3D grids are Euclidean data structures comprised of uniformly spaced 3D elements, termed voxels. These grids have distinct geometric properties: each voxel has a consistent neighborhood structure, making it indistinguishable from other voxels in terms of structure, and the vertices maintain a fixed order determined by their spatial dimensions. Owing to the Euclidean nature of the 3D grid input, 3D CNNs are conventionally employed to encode such data and to handle subsequent tasks.
\item The 3D surface of a protein is the exterior layer of the protein's atoms and is pivotal in protein-ligand interactions. Each point on this protein surface can be distinguished by its associated chemical (e.g., hydrophobic, electrostatic) and geometric (e.g., local shape, curvature) attributes. Protein surfaces can be represented as meshes, polygons that demarcate the surface's contours, or point clouds, sets of nodes that specify the surface's position at a particular resolution level.
\item 3D graphs are prevalently utilized to describe protein structural data, wherein atoms serve as nodes and covalent bonds as edges. Edges can also be formed by linking the k-nearest neighbors. Geometric GNNs \cite{satorras2021n, jing2021learning} are adept at processing protein 3D graphs.
\end{itemize}

\begin{figure}[t]
\centering
\begin{subfigure}[2D molecular graph]
{\includegraphics[width=.47\linewidth]{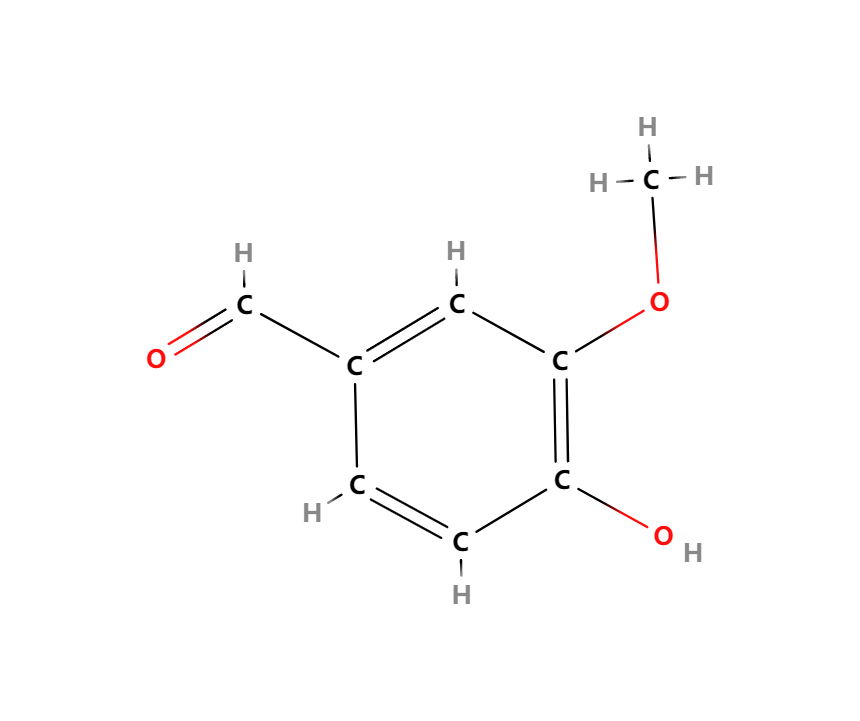} \label{fig:molecule_feature_example_2D}}
\end{subfigure}
\begin{subfigure}[3D molecular graph]
{\includegraphics[width=.47\linewidth]{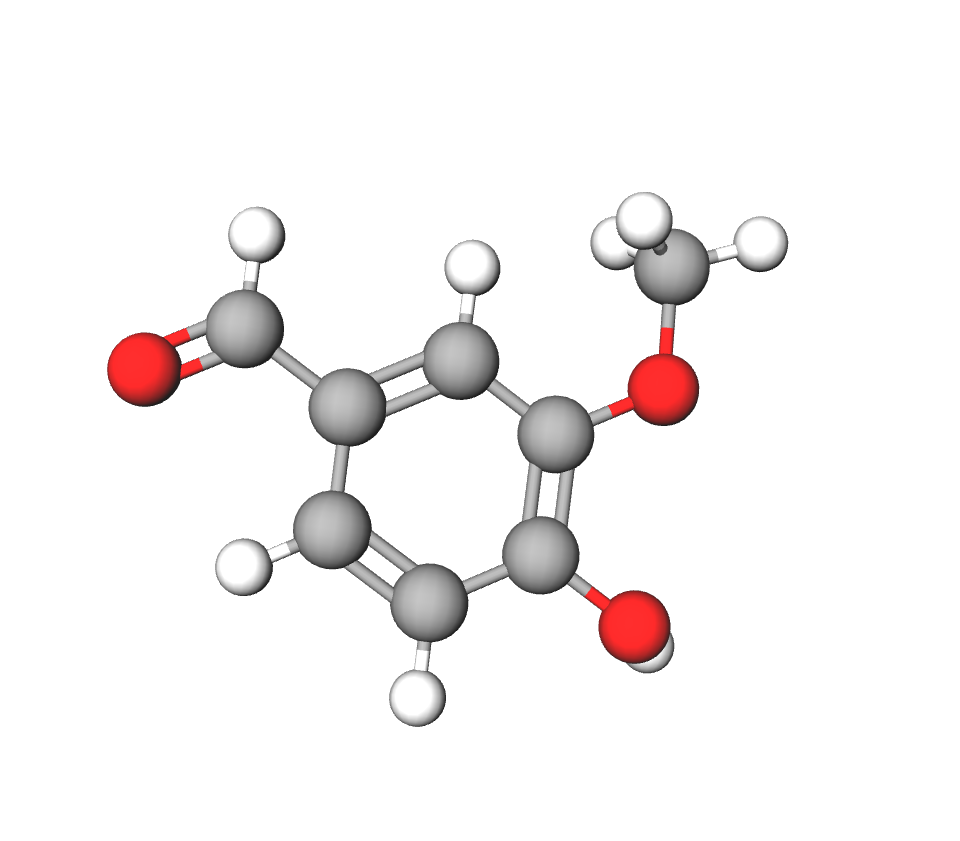} \label{fig:molecule_feature_example_3D}}
\end{subfigure}
\caption[Caption for LOF]{
    Representing molecules as (a) 2D graphs and (b) 3D graphs. 
    }
\label{fig:molecule_feature_examples}
\end{figure}

As for the ligand molecules, their representations vary from 1D strings, e.g., simplified molecular-input line-entry system (SMILES)~\cite{weininger1988smiles} to 2D and 3D graphs \cite{du2022molgensurvey} where nodes represent atoms and edges represent bonds (Figure~\ref{fig:molecule_feature_examples}). 

\subsection{Symmetries}
\label{symmetry}
Incorporating the symmetry priors into neural network architectures as inductive bias is an effective strategy to build generalizable models \cite{bronstein2017geometric, atz2021geometric}. 
The
main symmetry groups in protein-ligand systems include the Euclidean group E(3), the particular Euclidean group SE(3), and the permutation group \cite{atz2021geometric}.
Both E(3) and SE(3) include rotation and translation transformations in the 3D Euclidean space. 
E(3) further covers the reflection operation. 
These transformations are essential for geometric deep learning because the output should obey the underlying physics rules that predicted properties of proteins/ligands should not change under the transformation of coordinate systems and atom orders.
SE(3) is applicable when a neural network aims to differentiate chiral systems \cite{adams2021learning}, which are systems that are not superimposable on their mirror images, much like left and right human hands. In chemistry and biology, chiral molecules can exhibit unique properties, e.g., a drug might be therapeutic, while its mirror image might be harmful or inactive.
Formally, let $\mathcal{T}$ denote the transformation, $f$ be the neural network, and $\vx$ be the input data. The output of the neural network $f(\vx)$ can transform equivariantly or invariantly with respect to $\mathcal{T}$ if satisfy the following constraints:
\begin{itemize}
\item \textbf{Equivariance:} $f(\vx)$ is equivariant to a transformation $\mathcal{T}$
if the transformation of input $\vx$ commutes with the transformation of $f(\vx)$ via a transformation $\mathcal{T}'$ of the
same symmetry group, i.e, $f(\mathcal{T}(\vx)) = \mathcal{T}'f(\vx)$. Such symmetry is important as the predicted vector outputs (e.g., forces, coordinates) should not depend on the choice of coordinate systems.
\item \textbf{Invariance:} Invariance is a special case of equivariance where $f(\vx)$ is invariant to $\mathcal{T}$ if $\mathcal{T}'$ is the identity transformation: $f(\mathcal{T}(\vx)) = \mathcal{T}'f(\vx) = f(\vx)$. Such symmetry prior is important as the predicted scalar properties of a certain molecule/protein (e.g., energies) should be the same under the transformation of coordinate systems.
\end{itemize}

\subsection{Predictive Methods}
Next we summarize the main predictive methods for predictive tasks, i.e., binding site and affinity predictions. These methods are also widely used as the structure encoding backbones for generative models. 

\subsubsection{Convolutional Neural Networks (CNNs)} CNNs are widely used for image processing where the input elements, i.e., pixels, are arranged spatially. CNNs rely on the shared-weight architecture of convolution kernels or filters that slide along input features and provide translation-equivariant outputs. Convolution kernels and filters can vary with data structure. For example, for the 3D surface data, MaSIF \cite{gainza2020deciphering} defines geodesic convolutional layers with the geodesic polar coordinates. For the 3D grid data, 3D CNNs are widely used \cite{jimenez2017deepsite, kandel2021puresnet}.

\subsubsection{Graph Neural Networks (GNNs)} GNNs
are widely used to model relational data. Most GNNs follow a message-passing paradigm. Let the $h_i$ be the node feature of the $i$-th node in the graph and $e_{ij}$ be the edge feature for the optional edge connecting node $i$ and $j$. GNNs iteratively conduct message computation and
neighborhood aggregation operations for each node (or edge). Generally, we have:
\begin{align}
    m_{ij} &= \psi_m(h_i, h_j, e_{ij}),\\
    h_i' &= \psi_h(\{m_{ij}\}_{j\in\mathcal{N}(i)}, h_i),
\end{align}
where $\mathcal{N}(i)$ denote the set of neighbors of node $i$, $h_i'$ is the updated node feature, and $\psi_m,\psi_h$ are learnable functions. 

For the 3D structural data, each node in the 3D graph has scalar features and contains 3D coordinates. Equivariant graph neural networks are proposed to incorporate geometric symmetry into model building \cite{han2022geometrically}. Let $\bm v_i \in \mathbb{R}^3$ denote the 3D coordinate, we have:
\begin{align}
    m_{ij} &= \psi_m(\bm v_i, \bm v_j, h_i, h_j, e_{ij}),\\
    \bm m_{ij} &= \psi_{\bm m}(\bm v_i, \bm v_j, h_i, h_j, e_{ij}),\\
    h_i' &= \psi_h(\{m_{ij}\}_{j\in\mathcal{N}(i)}, h_i),\\
    \bm v_i' &= \psi_{\bm v}(\{\bm m_{ij}\}_{j\in\mathcal{N}(i)}, \bm v_i),
\end{align}
where $m_{ij}$ and $\bm m_{ij}$ are scalar and vector messages. $\psi_m$ and $\psi_h$ are geometrically invariant functions while $\psi_{\bm m}$ and $\psi_{\bm v}$ are geometrically equivariant functions.
Compared with traditional 3D CNNs, geometrically equivariant GNNs do not require the voxelization of input data while still maintaining the desirable equivariance. Some representative equivariant GNNs are SchNet \cite{schutt2018schnet}, EGNN \cite{satorras2021n}, GVP \cite{jing2020learning}, DimeNet~\cite{klicpera2020directional,klicpera2020fast}, GMN \cite{huang2022equivariant}, SphereNet~\cite{liu2021spherical}, and ComENet \cite{wang2022comenet}.

\subsubsection{Graph Transformers}
Transformers were originally developed for sequential data \cite{devlin2018bert}. Transformer architectures were recently adapted to 2D and 3D graph data and achieved superior performance \cite{ying2021transformers, li2022synthesis, rong2020self, liao2022equiformer, luo2022one} than GNNs on node and graph classification tasks.
A transformer is composed of stacked transformer blocks, where each block consists of two layers: a self-attention layer followed by a feed-forward layer with normalizations (e.g., LayerNorm \cite{he2016deep}) and skip connections for both layers. For an input feature matrix $\textit{\textbf{H}}^{(l)}$, the $(l + 1)$-th self-attention block works as follows:
\begin{align}
    \textit{\textbf{A}}^{(l)} &= {\rm softmax}\left(\frac{\textit{\textbf{H}}^{(l)}\textit{\textbf{W}}^{(l)}_Q(\textit{\textbf{H}}^{(l)}\textit{\textbf{W}}^{(l)}_K)^\top}{\sqrt{d}} \right);\\
    \textit{\textbf{H}}^{(l+1)} &= \textit{\textbf{H}}^{(l)} + \sum_{i=1}^B \textit{\textbf{A}}^{(l)} \textit{\textbf{H}}^{(l)} \textit{\textbf{W}}^{(l)}_V \textit{\textbf{W}}^{(l)}_O,
\end{align}
where $\textit{\textbf{A}}$ is the attention matrix, $B$ is the total number of attention heads, $d$ is the dimension size, $\textit{\textbf{W}}^{(l)}_Q$, $\textit{\textbf{W}}^{(l)}_K$, and $\textit{\textbf{W}}^{(l)}_V$ are learnable transformation matrices at layer $l$. 

To generalize transformers to graphs, positional encodings are indispensable for encoding topological and geometric information \cite{ying2021transformers, luo2022one}. 
Popular positional encodings are based on shortest paths \cite{ying2021transformers}, PageRank \cite{zhang2022hierarchical}, and eigenvectors \cite{kreuzer2021rethinking}.
Positional encodings are necessary for graph transformers to consider graph connectivity--without positional encodings the models would ignore the strong inductive bias of edges and would attend to any pair of nodes.
Representative graph transformers include Graphormer \cite{ying2021transformers}, Transformer-M \cite{luo2022one}, and Equiformer \cite{liao2022equiformer}. For example, Transformer-M \cite{luo2022one} develops two separated
channels to encode both 2D and 3D structural information of molecules. The 2D
channel uses encodings based on atom degrees, shortest path distance, and 2D graph structure. In the 3D channel, Gaussian basis kernel functions \cite{scholkopf1997comparing} are used to encode 3D spatial distances between atoms. 

\subsection{Generative Methods} 
\label{sec:generative_models} 
\label{sec:method}
The following section provides an overview of generative methods applied to SBDD tasks, encompassing autoregressive models, flow models, variational autoencoders, diffusion probabilistic models, and flow matching models. Unlike predictive methods that focus on making accurate predictions from input data, generative methods aim to create new data samples that resemble the training distribution. In SBDD, generative tasks include binding pose generation, de novo ligand generation, protein pocket generation, and linker design. While certain tasks, such as binding pose generation, can be approached as predictive tasks, framing them as generative tasks allows for a more comprehensive representation of both aleatoric and epistemic uncertainties inherent in the process. 

\subsubsection{Autoregressive Models (ARs)}
A data point $\vx$ can be factorized into a set of components $\{x_0, x_1, \dots, x_{d-1}\}$, where $d$ is the number of components. These components can be pixels in images and nodes and edges in graphs. The components may have complicated underlying dependencies, making the direct generation of $\vx$ challenging.  With predefined or learned orders, the autoregressive models factorize the joint distribution of $\vx$ into the product of $d$ likelihoods as follows:
\begin{equation} \small{ \label{eq:autoregressive}
p(\vx) = \prod_{i=1}^d p(x_i|x_1, x_2, ..., x_{i-1}).
} \end{equation}
Autoregressive models sequentially generate $\vx$: In each step of a generative process, the next subcomponent is predicted based on the previously generated subcomponents. 

\subsubsection{Flow Models} \label{sec:flow}
A flow model aims to learn a parameterized invertible function between a data point $\textbf{x}$ and the latent variable $\vz$: $f: \vz\in \mathbb{R}^d \xrightarrow{} \vx \in \mathbb{R}^d$.
The latent distribution $p(\vz)$ is a predefined probability distribution, e.g., a Gaussian distribution. The data distribution $p(\vx)$ is unknown. But given a data point $\vx$, its likelihood can be computed with the change-of-variable theorem:
\begin{equation} \small{\label{eq:change_of_variable_theorem}
p(\vx) = p(\vz) \Big| \text{det} \Big( \frac{d f^{-1}(\vx)}{d \vx} \Big) \Big|,
} \end{equation}
where $\text{det}(\cdot)$ is the matrix determinant and $\frac{d f^{-1}(\vx)}{d \vx}$ is the Jacobian matrix.

In the sampling process, a latent variable $\vz$ is first sampled from a predefined latent distribution $p(\vz)$. Then the corresponding data point $\vx$ is obtained by performing a feedforward transformation $\vx = f(\vz)$. Therefore, $f$ must be differentiable, and the computation of ${\rm det}J$ should be tractable for training and sampling efficiency. A common choice is the affine coupling layers \cite{dinh2014nice, dinh2016density} where the computation of ${\rm det}J$ is very efficient because $J$ is an upper triangular matrix.
\subsubsection{Variational Autoencoders (VAEs)} 
Variational autoencoders (VAEs)~\cite{kingma2013auto} maximize the evidence lower bound (ELBO) of $p(\vx)$. VAEs introduce the latent variable $\vz$ to express the likelihood of $\vx$ as:
\begin{align} \small \label{eq:variation_autoencoder}
\log p(\vx)
& = \log \int_\vz p(\vz) p(\vx|\vz) d\vz\\
& \ge \mathbb{E}_{q(\vz|\vx)} \big[ \log p(\vx|\vz) \big] - D_{KL}(q(\vz|\vx) || p(\vz))\\
& \triangleq \text{ELBO}.
\end{align}
Here, $p(\vz)$ represents the prior distribution of $\vx$. For tractable calculation, parameterized encoder $q(\vz|\vx)$ is usually used to approximate $p(\vz|\vx)$, also known as the variational inference technique. The first term of ELBO is reconstruction loss, which quantifies the information loss of reconstructing $\vx$ from latent representations. The standard Gaussian prior $p(\vz) \sim \mathcal{N}(0, \textit{\textbf{I}})$ typically leads to mean-squared error (MSE) loss for the first term.
The second term in ELBO is the KL-divergence term, ensuring that our learned distribution $q(\vz|\vx)$ is similar to the true prior distribution.

\subsubsection{Diffusion Probabilistic Models} 
Diffusion models~\cite{sohl2015deep,ho2020denoising} are generative models inspired by non-equilibrium thermodynamics. A diffusion model defines a Markovian chain of random
diffusion process where each step adds noise to the data, and then it learns the reverse of this process via neural networks to reconstruct data points from noise distributions, e.g., isotropic Gaussian.

Let $\vx_0 \sim p(\vx)$ denote the input data point and $\vx_t$ for $t=1,\dots ,T$ indicate a series of noised representations with the same dimension as $\vx_0$. The forward diffusion process can be expressed as:
\begin{equation} \small{
\begin{aligned}
q(\vx_t | \vx_{t-1})
& = \mathcal{N}(\vx_{t}; \sqrt{1-\beta_t} \vx_{t-1}, \beta_t \textit{\textbf{I}}),
\end{aligned}
} \end{equation}
where $\beta_t \in (0, 1)$ controls the strength of Gaussian noise added in each step. A desirable property of the diffusion process is that a closed
form of the intermediate state can be obtained. Let $\alpha_t = 1-\beta_t$, and $\bar \alpha_t = \prod_{i=1}^t \alpha_i$, we have:
\begin{equation}\small{
q(\vx_t | \vx_{0})
= \mathcal{N}(\vx_{t}; \sqrt{\Bar{\alpha}_t} \vx_{0}, (1-\Bar{\alpha}_t) \textit{\textbf{I}}).
}\label{x0xt} \end{equation}
Another desirable feature of the reverse diffusion process, i.e., the denoising process, is that it can be computed in a closed form when conditioned on $\vx_{0}$ using Bayes theorem:
\begin{equation} \small{ \label{eq:diffusion_backward}
\small{
\begin{aligned}
p_\theta(\vx_{t-1}|\vx_t) 
& = \mathcal{N}(\vx_{t-1}; \bm{\mu}_{t}(\vx_0, \vx_t), \widetilde{\beta}_{t} \textit{\textbf{I}}),
\end{aligned}
}
} \end{equation}
where the parameters can be obtained analytically:
\begin{equation} \small{\begin{aligned}
\bm{\mu}_{t}(\vx_0, \vx_t)& =\frac{\sqrt{\bar{\alpha}_{t-1}}\beta_t}{1-\bar{\alpha}_{t}}\vx_{0}+\frac{\sqrt{\alpha_t}(1-\bar{\alpha}_{t-1})}{1-\bar{\alpha}_{t}}\vx_{t}, \\ \widetilde{\beta}_{t}& =\frac{1-\bar{\alpha}_{t-1}}{1-\bar{\alpha}_{t}}\beta_t.\end{aligned}
} \end{equation}
Similar to variational autoencoders, the objective of diffusion models is to maximize the variational lower bound of log-likelihood of $p(\vx)$ as:
\begin{equation*} \small{
\begin{aligned}
& - \log p(\vx)
 \le \underbrace{KL[q(\vx_T|\vx_0) || p_\theta(\vx_T)]}_{\rm prior \ loss \ \mathcal{L}_T} \\
& ~~~ + \sum_{t=2}^T \underbrace{KL[q(\vx_{t-1}|\vx_t,\vx_0) || p_\theta(\vx_{t-1}|\vx_t)]}_{\rm diffusion \ loss \ \mathcal{L}_{t-1}} - \underbrace{ \mathbb{E}_q [\log p_\theta(\vx_0|\vx_1)]}_{\rm reconstruction \ loss \ \mathcal{L}_0}.
\end{aligned}
} \end{equation*}
Here, $\mathcal{L}_T$ is a constant and $\mathcal{L}_0$ can be estimated using an auxiliary model \cite{ho2020denoising,weng2021diffusion}. For $\{\mathcal{L}_{t-1}\}_{t=2}^T$, diffusion models adopt a neural network $\epsilon_\theta$ to predict the noise. More specifically, we can reparameterize Equation~\ref{x0xt} as $\vx_t = \sqrt{\Bar{\alpha}_t} \vx_{0}+ \sqrt{(1-\Bar{\alpha}_t)} \epsilon_t, \epsilon_t \sim \mathcal{N}(0, \textit{\textbf{I}})$.
The following training objective is widely adopted:
\begin{equation*} \small{
\mathcal{L} = \mathbb{E}_t \Big[\| \epsilon_t - \epsilon_\theta(\vx_t, t) \|^2 \Big].
} \end{equation*}

\subsubsection{Flow Matching Models}

Flow Matching (FM) \cite{lipman2022flow, albergo2022building, albergo2023stochastic} models are a class of generative models that aim to learn continuous transformations between simple base distributions and complex data distributions. Unlike diffusion models, which rely on stochastic processes, FM models utilize deterministic flows parameterized by neural networks to map between distributions, and have shown better performance and efficiency than diffusion models on a series of biomolecular tasks \cite{bose2023se, yim2023fast, song2024equivariant}.

In FM, the transformation from a base distribution \( p_0(\vx) \) to a target data distribution \( p_1(\vx) \) is defined through a time-dependent vector field \( \vu_t(\vx) \). This vector field governs the evolution of data points over time \( t \in [0,1] \) via the ordinary differential equation (ODE):
\begin{equation} \small{
\frac{d\vx}{dt} = \vu_t(\vx), \quad \vx_0 \sim p_0(\vx), \quad \vx_1 \sim p_1(\vx).
} \end{equation}
The goal is to learn \( \vu_t(\vx) \) such that integrating this ODE transports samples from \( p_0 \) to \( p_1 \).

Training involves minimizing the discrepancy between the learned vector field \( \vu_t(\vx) \) and a target vector field \( \vv_t(\vx) \) that defines a desired probability path between \( p_0 \) and \( p_1 \). This is achieved by optimizing the following objective:
\begin{equation} \small{
\mathcal{L}_{\text{FM}} = \mathbb{E}_{t \sim \mathcal{U}[0,1], \vx \sim p_t(\vx)} \left[ \| \vu_t(\vx) - \vv_t(\vx) \|^2 \right],
} \end{equation}
where \( p_t(\vx) \) represents the intermediate distribution at time \( t \). The target vector field \( \vv_t(\vx) \) can be derived from various probability paths, including those based on optimal transport or diffusion processes.

Flow Matching models offer several advantages over traditional diffusion models: First, the use of ODEs allows for deterministic generation of samples, potentially reducing the variance associated with stochastic methods; Second, FM models can incorporate different probability paths, such as those derived from optimal transport theory, leading to more efficient training and sampling processes; Finally, by leveraging continuous normalizing flows, FM models can be trained at scale, facilitating their application to high-dimensional data.


\begin{table*}
  \centering
\caption{Summary of Datasets for SBDD
}
\begin{adjustbox}{width=1\textwidth}
\renewcommand{\arraystretch}{1.1}
\label{tab:datasets}
\begin{tabular}{lccccc}
\toprule
Category & Dataset & Size & Description \\
\midrule
Protein & \textbf{Protein Data Bank} \cite{berman2000protein}  & $\sim$223k & Experimentally determined biomolecule structures. \\
 & \textbf{AlphaFold DB} \cite{berman2000protein}  & $\sim$200m & Alphafold predicted protein structures for proteins in UniProt.\\
\midrule
Small Molecule & \textbf{ZINC250k}~\cite{sterling2015zinc} & 250k & Commercially available small molecules selected based on drug-likeness and structural diversity. \\
& \noindent\textbf{CASF} \cite{su2018comparative} & & Experimentally determined 3D structures of small molecules. \\
& \noindent\textbf{GEOM} \cite{axelrod2022geom} & $\sim$37m & DFT generated molecular conformations for over 450k molecules.\\
& \noindent\textbf{PROTAC-DB}~\cite{Ge2024PROTACDB3A} &  6,111 & PROTAC molecules manually collected from scientific literature. \\
\midrule
Binding Complex & \textbf{PDBBind} \cite{liu2017forging} & 19,443 & Comprehensive collection of experimental binding complexes structures and binding affinity. \\
& \textbf{scPDB~\cite{Desaphy2015-scpdb}} & 16,034 & Annotated Database of Druggable Binding Sites from the PDB. \\
& \noindent\textbf{CASF} \cite{su2018comparative} & 285 &  Complexes with high-quality crystal structures and reliable binding constants in PDBBind. \\
 & \textbf{CrossDocked} \cite{francoeur2020three} & $\sim$22.5m & Constructed by cross-docking the ligands into multiple similar binding pockets across the PDB. \\
& \noindent\textbf{Binding MOAD} \cite{hu2005binding} & 41,409 & Experimentally determined binding complexes structures and measured binding affinity. \\
& \noindent\textbf{BioLiP~\cite{Yang2013-gp-biolip}} & 903,398 & Semi-manually curated database for biologically relevant ligand-protein binding interactions.\\
& \textbf{CSAR-HiQ}  \cite{Dunbar2013CSARDS} & 343 & High-quality experimental binding complexes structures and binding affinity. \\
& \textbf{PoseBusters} \cite{Buttenschoen2023PoseBustersAD} & 308 &  Selected recent high-quality protein-ligand complexes which contain
drug-like molecules.\\
& \textbf{DockGen~\cite{Corso2023-da-dockgen}} & 330 & A benchmark based on the ligand-binding domains of proteins focusing on generalizability.\\
& \textbf{APObind~\cite{Aggarwal2021APObindAD}}  &  10,599 & A dataset provides apo protein structures for holo structures present in the PDBbind. \\

& \textbf{PLINDER} \cite{Durairaj2024PLINDERTP} & 449,383 & A  large and comprehensively annotated protein-ligand interaction dataset. \\
& \textbf{COACH420} \cite{Yang2013-COACH420} & 420 & Single chain protein structures and a mix of drug targets and naturally occurring
ligands \\
& \textbf{HOLO4K} \cite{Schmidtke2010-holo4k} & 4,009 & A large protein-ligand complexes dataset containing larger multi-chain protein structures. \\
\bottomrule
\end{tabular}
\renewcommand{\arraystretch}{1}
\end{adjustbox}
\end{table*}

\begin{table*}
  \centering
\caption{SBDD Terms Glossary
}
\begin{adjustbox}{width=1\textwidth}
\renewcommand{\arraystretch}{1.1}
\label{tab:glossary}
\begin{tabular}{lccccc}
\toprule
 & Term & Definition \\
\midrule
 & \textbf{Protein} & A macromolecule made up of one or more chains of amino acids linked by peptide bonds. \\
& \textbf{Ligand} & A molecule that binds to another (usually larger) molecule\\
& \textbf{Linker} & A chemical matter that connects two segments of a larger molecule or functional groups. \\
& \textbf{Binding Site/Porcket} & A region on a macromolecule such as a protein that binds to another molecule with specificity. \\
& \textbf{Binding Complex} & A stable assembly formed by two or more molecules, such as a protein and its ligand.\\
& \textbf{Protein Apo Structure} & The structure of protein on its own, before it interacts with other ligand.\\
& \textbf{Protein Holo Structure} & The structure of protein binding with other ligand.\\
& \textbf{Molecule fragment} & Substructure of a molecule that represents a smaller portion of the overall molecular structure.\\
& \textbf{PROTAC} & A class of heterobifunctional molecule designed to degrade specific target proteins.\\
& \textbf{Molecular Docking} & A computational method that predicts the optimal interaction between two molecules.\\
& \textbf{Rigid Docking} & Docking approach where both molecules are treated as rigid bodies without any conformational changes.\\
& \textbf{Flexible Docking} & A docking approach that allows for flexibility and conformational changes in the structure of the molecules.\\
& \textbf{Site-specific Docking} & Docking approach that focuses on a known binding site on the target molecule.\\
& \textbf{Blind Docking} & A docking approach that does not assume a specific binding site.\\
& \textbf{Search Space} & The defined region on a molecular target where the docking algorithm explores potential binding interactions.\\
& \textbf{Binding Affinity} & A measure of the strength of the interaction between two molecules, e.g. protein and ligand.\\
& \textbf{Co-crystal Structure} & The experimentally determined 3D structure of a complex formed between two or more molecules.\\

\midrule
 & \textbf{RL (Reinforcement learning)} & An agent learns to optimize decisions by interacting with an environment and receiving rewards or penalties.\\
& \textbf{GA (Genetic Algorithm)} & An optimization method that uses selection, crossover, and mutation to evolve solutions.\\
& \textbf{MCTS (Monte Carlo Tree Search)} & A decision-making algorithm useing random sampling to explore possible moves and optimize outcomes.\\
& \textbf{VD (Variational Dropout)} & A regularization method that treats dropout rates as random variables to prevent overfitting in neural networks.\\
& \textbf{Iterative Update} & Iteratively refine the molecule structure based on current context.\\
& \textbf{Dist. Pred.} & Predicting the Euclidean distances between atoms or molecular fragments to determine the 3D molecule structure.\\
& \textbf{Flow Matching} & A generative method that generalizes diffusion, learning flow between arbitrary start and end distributions.\\

\bottomrule
\end{tabular}
\renewcommand{\arraystretch}{1}
\end{adjustbox}
\end{table*}

\subsection{Other Methods}

Apart from the generative methods previously discussed, several alternative techniques such as Reinforcement Learning (RL) \cite{chen20233d}, Genetic Algorithm (GA) \cite{fu2022reinforced}, and Monte Carlo Tree Search (MCTS) \cite{li2021structure, li2022synthesis} are utilized to probe the chemical space for properties of interest. Additionally, drawing inspiration from molecular dynamics and fragment-based drug design, innovative methods grounded on virtual dynamics (VD) \cite{lu2023d} and fragment-based molecule generation (Fragment) \cite{zhang2023molecule} have been introduced. For instance, in generating 3D molecules suited to a target protein, VD-Gen \cite{lu2023d} initializes numerous virtual particles within the pocket. These particles are then iteratively adjusted to mirror the spatial arrangement of molecular atoms. A 3D molecule is subsequently derived from the stabilized virtual particles. In contrast, in fragment-based molecule generation, a motif vocabulary is initially established by isolating prevalent molecular fragments (referred to as motifs) from the dataset. During the generative phase, molecules are generated by autoregressively appending new fragments, ensuring the resulting local structures maintain realism.

\begin{table*}[]
\caption{Summary of structure-based drug design models with geometric deep learning reviewed in this paper.}
\label{tab:sbdd}
\centering
\begin{tabular}{l l l l l l l}
\toprule
Task & Model & Input & Output & Method \\ 
\midrule
\multirow{15}{*}{Binding site prediction} 
& MaSIF~\cite{gainza2020deciphering} & Protein Surface & Binding Site Probability & CNN\\
& dMaSIF~\cite{sverrisson2021fast} & Protein Surface & Binding Site Probability & CNN\\
\cmidrule(lr){2-5}
& PeSTo~\cite{krapp2023pesto} & Protein 3D-Graph & Binding Site Probability & Transformer \\
\cmidrule(lr){2-5}
& ScanNet~\cite{tubiana2022scannet} & Protein 3D-Graph & Binding Site Probability & GNN \\
& PocketMiner~\cite{meller2023predicting} & Protein 3D-Graph & Binding Site Probability & GNN \\
& PIPGCN~\cite{fout2017protein} & Protein 3D-Graph & Binding Site Probability & GNN \\
& EquiPocket~\cite{zhang2023equipocket} & Protein 3D-Graph & Binding Site Probability & GNN \\
& NodeCoder~\cite{abdollahi2023nodecoder} & Protein 3D-Graph & Binding Site Probability & GNN \\
& VN-EGNN~\cite{sestak2023vnegnn} & Protein 3D-Graph & Binding Site Probability & GNN \\
\cmidrule(lr){2-5}
& DeepSite~\cite{jimenez2017deepsite} & Protein Grid & Binding Site Probability & 3DCNN \\
& Kalasanty~\cite{Stepniewska-Dziubinska2020-Kalasanty} & Protein Grid & Binding Site Probability & 3DCNN \\
& DeepSurf~\cite{Mylonas2021-DeepSurf} & Protein Grid & Binding Site Probability & 3DCNN \\
& DeepPocket~\cite{Aggarwal2022-DeepPocket} & Protein Grid & Binding Site Probability & 3DCNN \\
& PUResNet~\cite{kandel2021puresnet} & Protein Grid & Binding Site Probability & 3DCNN \\
& DSDP~\cite{Huang2023DSDPAB} & Protein Grid & Binding Site Probability & 3DCNN \\
\midrule 
\multirow{28}{*}{Binding pose generation} 
& EquiBind~\cite{stark2022equibind} & Ligand 2D-Graph+Protein 3D-Graph & Complex 3D-Graph & Keypoint Align\\
\cmidrule(lr){2-5}
& TankBind~\cite{lu2022tankbind} & Ligand 2D-Graph+Protein 3D-Graph & Complex 3D-Graph & Dist. Pred.\\
& EDM-Dock~\cite{masters2023deep} & Ligand 2D-Graph+Protein 3D-Graph & Complex 3D-Graph & Dist. Pred.\\
\cmidrule(lr){2-5}
& DPL~\cite{ProfileShuya2022structure} & Ligand 2D-Graph+Protein Sequence & Complex 3D-Graph & Diffusion\\
& DiffDock~\cite{corso2022diffdock} & Ligand 2D-Graph+Protein 3D-Graph & Complex 3D-Graph & Diffusion\\
& NeuralPLexer~\cite{qiao2022state} & Ligand 2D-Graph+Protein Sequence & Complex 3D-Graph & Diffusion\\
& DynamicBind~\cite{lu2023dynamicbind}& Ligand 2D-Graph+Protein 3D-Graph & Complex 3D-Graph & Diffusion\\
& DiffDock-Pocket~\cite{plainer2023diffdockpocket} & Ligand 2D-Graph+Protein 3D-Graph & Complex 3D-Graph & Diffusion\\
& DiffBindFR~\cite{Zhu2023DiffBindFRAS} & Ligand 2D-Graph+Protein 3D-Graph & Complex 3D-Graph & Diffusion\\
& DiffDock-L~\cite{corso2024deep} & Ligand 2D-Graph+Protein 3D-Graph & Complex 3D-Graph & Diffusion\\
& SurfDock~\cite{Cao2023SurfDockIA} & Ligand 2D-Graph+Protein 3D-Graph & Complex 3D-Graph & Diffusion\\
& PackDock~\cite{Zhang2024PackDockAD} & Ligand 2D-Graph+Protein 3D-Graph & Complex 3D-Graph & Diffusion\\
& Re-Dock~\cite{Huang2024ReDockTF} & Ligand 2D-Graph+Protein 3D-Graph & Complex 3D-Graph & Diffusion\\
& HelixDock~\cite{Liu2023PreTrainingOL} & Ligand 2D-Graph+Protein 3D-Graph & Complex 3D-Graph & Diffusion\\
& RosettaFold All-Atom~\cite{krishna2023generalized} & Ligand 2D-Graph+Protein Sequence & Complex 3D-Graph & Diffusion\\
& AlphaFold3~\cite{Abramson2024AccurateSP} & Ligand 2D-Graph+Protein Sequence & Complex 3D-Graph & Diffusion\\
& Chai-1~\cite{chai1} & Ligand 2D-Graph+Protein Sequence & Complex 3D-Graph & Diffusion\\
& Umol~\cite{Bryant2023StructurePO} & Ligand 2D-Graph+Protein Sequence & Complex 3D-Graph & Diffusion\\
\cmidrule(lr){2-5}
& HarmonicFlow~\cite{Stark2024-yc-harmonicflow} & Ligand 2D-Graph+Protein 3D-Graph & Complex 3D-Graph & Flow Matching\\
\cmidrule(lr){2-5}
& E3Bind~\cite{zhang2022e3bind} & Ligand 2D-Graph+Protein 3D-Graph & Complex 3D-Graph & Iterative Update\\
& DeepRMSD~\cite{wang2023fully} & Ligand 2D-Graph+Protein 3D-Graph & Complex 3D-Graph & Iterative Update\\
& 3T~\cite{mailoa2023protein} & Ligand 2D-Graph+Protein 3D-Graph & Complex 3D-Graph & Iterative Update\\
& UniMol~\cite{zhou2023unimol} & Ligand 2D-Graph+Protein 3D-Graph & Complex 3D-Graph & Iterative Update\\
& FABind~\cite{Pei2023FABindFA} & Ligand 2D-Graph+Protein 3D-Graph & Complex 3D-Graph & Iterative Update\\
& FlexPose~\cite{Dong2023EquivariantFM} & Ligand 2D-Graph+Protein 3D-Graph & Complex 3D-Graph & Iterative Update\\
& KarmaDock~\cite{Zhang2023EfficientAA} & Ligand 2D-Graph+Protein 3D-Graph & Complex 3D-Graph & Iterative Update\\
& CarsiDock~\cite{Cai2023CarsiDockAD} & Ligand 2D-Graph+Protein 3D-Graph & Complex 3D-Graph & Iterative Update\\
& FABind+~\cite{Gao2024FABindEM} & Ligand 2D-Graph+Protein 3D-Graph & Complex 3D-Graph & Iterative Update\\
& UniMol-V2~\cite{Alcaide2024UniMolDV} & Ligand 2D-Graph+Protein 3D-Graph & Complex 3D-Graph & Iterative Update\\
\midrule
\multirow{20}{*}{Binding affinity prediction} 
& SIGN~\cite{10.1145/3447548.3467311} & Complex 3D-Graph & Binding Affinity & GNN \\ 
& PIGNet~\cite{moon2022pignet} & Complex 3D-Graph & Binding Affinity & GNN \\
& HOLOPROT~\cite{somnath2021multi} & Complex 3D-Graph+Surface & Binding Affinity & GNN \\
& PLIG~\cite{moesser2022protein} & Complex 3D-Graph & Binding Affinity & GNN \\
& PaxNet~\cite{zhang2022efficient}& Complex 3D-Graph & Binding Affinity & GNN \\
& IGN~\cite{jiang2021interactiongraphnet} & Complex 3D-Graph & Binding Affinity & GNN \\
& GIGN~\cite{yang2023geometric} & Complex 3D-Graph & Binding Affinity & GNN \\
& GraphscoreDTA~\cite{Wang2023GraphscoreDTAOG} & Complex 3D-Graph & Binding Affinity & GNN \\
& PLANET~\cite{Zhang2023PLANETAM} & Complex 3D-Graph & Binding Affinity & GNN \\
& DOX\_BDW~\cite{Liu2023DOXBDWIS} & Complex 3D-Graph & Binding Affinity & GNN \\
& MBP~\cite{yan2023multi}& Complex 3D-Graph & Binding Affinity & GNN \\
& GIANT~\cite{Li2024GIaNtPB} & Complex 3D-Graph & Binding Affinity & GNN \\
\cmidrule(lr){2-5} 
& GET~\cite{Kong2023GeneralistET} & Complex 3D-Graph & Binding Affinity & Transformer \\
& BindNet~\cite{Feng2023ProteinligandBR} & Complex 3D-Graph & Binding Affinity & Transformer \\
\cmidrule(lr){2-5}
& Pafnucy~\cite{StepniewskaDziubinska2018DevelopmentAE} & Complex 3D-Grid & Binding Affinity & CNN \\
& DeepAtom~\cite{Li2019DeepAtomAF} & Complex 3D-Grid & Binding Affinity & CNN \\
& RoseNet ~\cite{HassanHarrirou2020RosENetIB} & Complex 3D-Grid & Binding Affinity & CNN \\
& IEConv~\cite{hermosilla2020intrinsic}& Complex 3D-Graph& Binding Affinity & CNN \\
& SGCNN~\cite{lu2023improving}& Complex 3D-Graph& Binding Affinity & CNN \\
\cmidrule(lr){2-5}
& Fusion~\cite{Jones2020ImprovedPB}& Complex 3D-Grid+3D-Graph & Binding Affinity & CNN + GNN \\
\bottomrule
\end{tabular}
\end{table*}
\setcounter{table}{2}
\begin{table*}
\caption{Summary of structure-based drug design models with geometric deep learning reviewed in this paper (continued).}
\centering
\begin{tabular}{l l l l l l l}
\toprule
Task & Model & Input & Output & Method \\ 
\midrule
\multirow{30}{*}{\emph{de novo} ligand generation} 
& AutoGrow~\cite{spiegel2020autogrow4} & Pocket 3D-Graph & Ligand 3D-Graph & GA \\
& RGA~\cite{fu2022reinforced}& Pocket 3D-Graph & Ligand 3D-Graph & GA \\
\cmidrule(lr){2-5}
& liGAN~\cite{masuda2020generating} & Pocket Grid & Ligand 3D-Graph & VAE \\
& RELATION~\cite{wang2022relation} & Pocket Grid & Ligand Smiles & VAE \\
& SQUID~\cite{adams2022equivariant} & Target Shape Point Cloud & Ligand 3D-Graph & VAE+Fragment \\
\cmidrule(lr){2-5}
& DESERT~\cite{long2022zero} & Pocket Grid & Ligand 3D-Graph & AR \\
& DeepLigBuilder~\cite{li2021structure} & Pocket 3D-Graph & Ligand 3D-Graph & MCTS+RL \\
& DeepLigBuilder+ \cite{li2022synthesis} & Pocket 3D-Graph &Ligand 3D-Graph & MCTS+RL\\
\cmidrule(lr){2-5}
& 3DSBDD~\cite{luo20213d} & Pocket 3D-Graph & Ligand 3D-Graph & AR \\
& Pocket2Mol~\cite{peng2022pocket2mol} & Pocket 3D-Graph & Ligand 3D-Graph & AR \\
& PrefixMol~\cite{gao2023prefixmol} & Pocket 3D-Graph & Ligand Smiles & AR\\
& ResGen~\cite{zhang2023resgen} & Pocket 3D-Graph & Ligand Smiles & AR\\
& FLAG~\cite{zhang2023molecule} & Pocket 3D-Graph & Ligand 3D-Graph & AR+Fragment \\
& DrugGPS~\cite{zhang2023drug} & Pocket 3D-Graph & Ligand 3D-Graph & AR+Fragment \\
& Lingo3DMol~\cite{wang2023lingo3dmol} & Pocket 3D-Graph & Ligand 3D-Graph & AR+Fragment \\
\cmidrule(lr){2-5}
& GraphBP~\cite{liu2022graphbp} & Pocket 3D-Graph & Ligand 3D-Graph & Flow \\
& MolCode~\cite{zhang2023equivariant} & Pocket 3D-Graph & Ligand 3D-Graph & Flow \\
& GraphVF~\cite{sun2023graphvf} & Pocket 3D-Graph & Ligand 3D-Graph & Flow+Fragment \\
& SENF ~\cite{rozenberg2023structurebased} & Pocket 3D-Graph & Ligand 3D-Graph & Flow \\
\cmidrule(lr){2-5}
& DiffSBDD~\cite{schneuing2022structure}& Pocket 3D-Graph & Ligand 3D-Graph & Diffusion \\
& TargetDiff~\cite{guan2023d}& Pocket 3D-Graph & Ligand 3D-Graph & Diffusion \\
& DiffBP~\cite{lin2022diffbp}& Pocket 3D-Graph & Ligand 3D-Graph & Diffusion \\
& DecompDiff~\cite{guan2023decompdiff}& Pocket 3D-Graph & Ligand 3D-Graph & Diffusion \\
& IRDiff~\cite{huanginteraction}& Pocket 3D-Graph & Ligand 3D-Graph & Diffusion \\
& BindDM~\cite{huang2024binding}& Pocket 3D-Graph & Ligand 3D-Graph & Diffusion \\
& PMDM~\cite{huang2024dual}& Pocket 3D-Graph & Ligand 3D-Graph & Diffusion \\
& ShapeMol \cite{chen2023shape}& Target Shape Point Cloud & Ligand 3D-Graph & Diffusion \\

\cmidrule(lr){2-5}
& FlexSBDD~\cite{zhang2024flexsbdd}& Pocket 3D-Graph & Ligand 3D-Graph & Flow Matching \\
\cmidrule(lr){2-5}
& VD-Gen~\cite{lu2023d}& Pocket 3D-Graph & Ligand 3D-Graph & VD \\
\midrule
\multirow{6}{*}{Protein Pocket Generation}
& FAIR~\cite{zhang2023full} & Ligand 3D-Graph & Pocket 3D-Graph & Iterative Update\\
& PocketGen~\cite{zhang2024efficient} & Ligand 3D-Graph & Pocket 3D-Graph & Iterative Update\\
\cmidrule(lr){2-5}
&RFDiffusion~\cite{watson2023novo} & Ligand 3D-Graph & Pocket 3D-Graph & Diffusion\\
& RFDiffusionAA~\cite{krishna2024generalized} & Ligand 3D-Graph & Pocket 3D-Graph & Diffusion\\
\cmidrule(lr){2-5}
& PockeFlow~\cite{zhang2024generalized} & Ligand 3D-Graph & Pocket 3D-Graph & Flow Matching\\
\midrule
\multirow{10}{*}{Linker design}
& DeLinker~\cite{imrie2020deep} & Pocket 3D-Graph+Ligand Fragments & Ligand 2D-Graph & VAE\\
& 3Dlinker~\cite{huang20223dlinker}& Pocket 3D-Graph+Ligand Fragments & Ligand 3D-Graph & VAE\\
& DEVELOP \cite{imrie2021deep} & Pocket 3D-Graph+Ligand Fragments & Ligand 3D-Graph & VAE\\
\cmidrule(lr){2-5}
& Link-INVENT~\cite{guo2023link}& Pocket 3D-Graph+Ligand Fragments & Ligand 2D-Graph & RL\\
& PROTAC-INVENT~\cite{chen20233d}& Pocket 3D-Graph+Ligand Fragments & Ligand 3D-Graph & RL\\
& PROTAC-RL~\cite{Zheng2022AcceleratedRP} & Pocket 3D-Graph+Ligand Fragments & Ligand 3D-Graph & RL\\
\cmidrule(lr){2-5}
& DiffLinker~\cite{igashov2022equivariant}& Pocket 3D-Graph+Ligand Fragments & Ligand 3D-Graph & Diffusion \\
& LinkerNet~\cite{Guan2023LinkerNetFP} & Pocket 3D-Graph+Ligand Fragments & Ligand 3D-Graph & Diffusion \\
& DiffPROTACS~\cite{Li2024DiffPROTACsIA} & Pocket 3D-Graph+Ligand Fragments & Ligand 3D-Graph & Diffusion \\
\bottomrule
\end{tabular}
\end{table*}

\subsection{Datasets} 
\label{sec:task} 

Machine learning models for structure-based drug design are trained on protein and small molecule structure datasets (Table \ref{tab:datasets}). Data generated by in vitro experiments are generally considered ground truth data. Protein Data Bank (PDB) is a global repository consisting of experimentally determined three-dimensional structures of biomolecules. These 3D structures has atom-level resolution and are obtained through experimental techniques like X-ray crystallography, NMR spectroscopy, and cryo-electron microscopy.

However, due to the complexity and cost of experiments, the size of the experimental datasets are often small. Many datasets have been crafted through in silico methods to mitigate the issue of insufficient structural data. Some structural datasets are generated using in silico methods from other data modalities. For example, the AlphaFold Database (AFDB) is created using Alphafold to generate the structures of $\sim$200 million protein sequences. In addition, data can also be expanded by leveraging existing experimental structural data. For example, the CrossDocked dataset is generated by computationally docking ligands into multiple similar binding pockets across the PDB. This significantly increased the available number of ligand-protein complexes by orders of magnitude.

Biomolecular datasets employ various representations tailored to specific research needs. For instance, ZINC250k comprises 250,000 molecules encoded as SMILES strings, capturing the 2D connectivity of atoms within each molecule.  In contrast, the GEOM dataset provides 3D structures of molecular conformations generated via density functional theory (DFT), offering both atomic connectivity and precise 3D coordinates for each atom, which is essential for studies requiring spatial molecular information.  Additionally, certain datasets are curated from existing biomolecular structures to address specific research objectives. An example is scPDB, a specialized database focusing on ligand binding sites within proteins. It is constructed by selecting proteins complexed with small synthetic or natural ligands from the Protein Data Bank (PDB), thereby facilitating studies related to protein-ligand interactions. 

\section{Structure-based Drug Design Tasks} 
\label{sec:task} 

\subsection{Binding Site Prediction} \label{sec:binding_site}

\subsubsection{Problem Formulation}

The protein surface encompasses the outermost regions of proteins, interacting with the environment. It is typically characterized as a continuous shape with added geometric and chemical attributes. Predicting binding sites based on these protein surface representations is fundamental for other SBDD tasks, including binding pose generation and \emph{de novo} ligand generation.
Formally, let's represent the target protein surface as $\mathcal{S}$ (for instance, in the form of a mesh or point cloud). The goal is to devise a predictive model $f(\mathcal{S})$ that determines the likelihood of each point on the surface being a binding site.

\subsubsection{Representative Methods}

\begin{figure}[t]
\centering
\includegraphics[width=.98\linewidth]{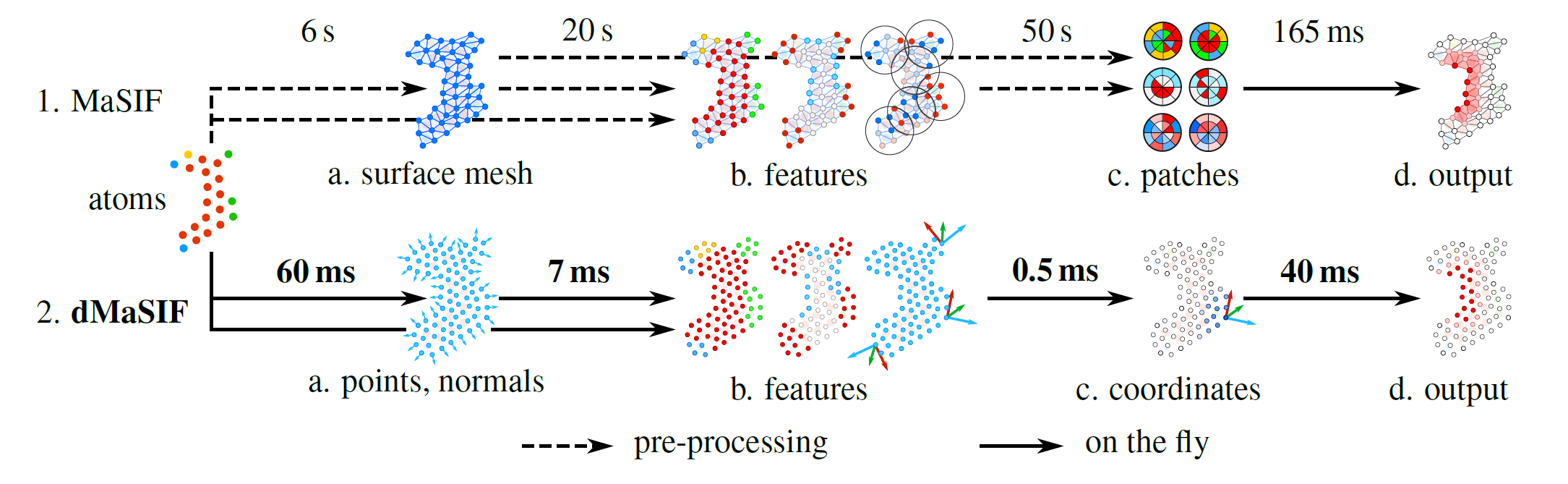}
\caption[]{Overview of MaSIF \cite{gainza2020deciphering} and dMaSIF~\cite{sverrisson2021fast} for binding site prediction. They have similar steps, and each step's average running time per protein is marked. MaSIF precomputes steps in a-c, whereas dMaSIF computes them on the fly and is 600 times faster than MaSIF. }
\label{fig:damasif}
\end{figure}

The Molecular Surface Interaction Fingerprinting (MaSIF) \cite{gainza2020deciphering} is a pioneering method that use
3D mesh-based geometric deep learning to predict protein interaction sites (first row in Figure~\ref{fig:damasif}). 
In MaSIF, the protein surface data is described in the geodesic space, where the distance between two points on the surface is measured by ``walking'' between the two points along the surface.
To encode the protein surface, MaSIF decomposes the surface into 
overlapping radial patches with a fixed geodesic radius. 
Each point within a patch is assigned an array of precomputed geometric (e.g., shape index and distance-dependent curvature) and 
chemical (e.g., hydropathy, continuum electrostatics, and free electrons/protons) features (Figure~\ref{fig:damasif}bc). MaSIF then learns to embed the surface patch’s input features into fingerprints for binding site prediction with convolutional neural networks (Figure~\ref{fig:damasif}d).

However, MaSIF \cite{gainza2020deciphering} is limited by the reliance on precomputed meshes, handcrafted features, and significant computation time. 
dMaSIF \cite{sverrisson2021fast} extend MaSIF and proposes an efficient end-to-end prediction framework based on 3D point cloud representations of protein. In Figure~\ref{fig:damasif}, it is shown that dMaSIF \cite{sverrisson2021fast} conducts all the computations on the fly and is 600 times faster than MaSIF \cite{gainza2020deciphering} while obtaining prediction results with a similar accuracy level.

Besides mesh representations, DeepSite \cite{jimenez2017deepsite} uses 3D grid to represents the protein molecules. Inspired from the computer vision perspective, the 3D protein image spans the bounding box of the target protein, and then discretized into a grid of $1 \times 1 \times 1$ $\AA$ sized grids. A 3D convolutional neural network (3DCNN) is trained on 3D protein images to predict the binding site. DeepSurf \cite{Mylonas2021-DeepSurf} uses the protein surface information by constructing local voxel grids around points on the protein surface and then trains 3DCNN to predict the binding site. However, these methods are not rotational invariant to the input protein structure. 

Some recent works model protein surfaces as 3D graphs and design GNN \cite{tubiana2022scannet} or Graph Transformer-based methods \cite{krapp2023pesto} for efficient and precise binding site prediction. For example, ScanNet \cite{tubiana2022scannet} builds representations of atoms and amino acids based on the spatial-chemical arrangement of protein and leverages GNN with specially designed filters for prediction. ScanNet predictions are local and invariant on Euclidean transformations. EquiPocket is an E(3)-equivariant GNN model for binding site prediction. EquiPocket explicitly incorporates surface geometric information of proteins, in which surface atoms comprise the major part of the binding pocket. PeSTo \cite{krapp2023pesto} is a rotation-equivariant transformer-based neural network that acts directly on protein atoms for the prediction of the binding site.

\subsubsection{Datasets}
\textbf{Protein Data Bank (PDB)} \cite{berman2000protein} contains 3D structural protein data obtained by X-ray crystallography, NMR spectroscopy, and cryo-electron microscopy. PDB provides the corresponding protein-ligand binding complex for proteins with known binding targets.

\noindent\textbf{scPDB \cite{Desaphy2015-scpdb}} is a comprehensive collection of ligandable binding sites of the Protein Data Bank. It includes all-atom information of the protein, ligand, binding site, and their binding mode. 

\noindent\textbf{COACH420~\cite{Yang2013-COACH420}} consists of 420 single chain protein structures and a mix of drug targets and naturally occurring ligands. 

\noindent\textbf{HOLO4K~\cite{Schmidtke2010-holo4k}} is a large protein-ligand complexes dataset, containing larger multi-chain protein structures.


\subsubsection{Evaluation Metrics}

\textbf{ROC-AUC} is widely used to evaluate the prediction of binding sites. The ROC-AUC assesses the model's accuracy in correctly classifying binding sites as a binary classification problem.

\noindent\textbf{DCC} is defined as the distance between the predicted binding site center and the true binding site center, while \textbf{DCA} is defined as the shortest distance between the predicted binding site center and any atom of the ligand. DCC and DCA are also widely used to used to benchmark the performance of binding site prediction models, providing insights into how closely the predicted sites align with actual binding regions. The benchmark performance are usually reported as
percentage of predictions with DCC/DCA less than a predefined threshold. 

\subsubsection{Benchmark Performance}
In Table~\ref{tab:binding site prediction benchmark}, we summarized the performance of selected binding site prediction methods on COACH420, HOLO4K, and PDBbind v2020 dataset. The performances are reported as the percentage of predictions with DCC/DCA less than 4~\AA. In general, GNN based methods such as VN-EGNN perform better than CNN based methods. 

\begin{table}
    \centering
    \caption{Summary of performance for representative binding site prediction method on different datasets. }
    
\begin{tabular}{c|c|c|c|c|c|c}
\toprule
\multirow{2}{*}{Method} & \multicolumn{2}{c|}{COACH420} & \multicolumn{2}{c|}{HOLO4K} & \multicolumn{2}{c}{PDBbind2020} \\
& DCC & DCA & DCC & DCA & DCC & DCA\\\midrule
\midrule 
Kalasanty~\cite{Stepniewska-Dziubinska2020-Kalasanty} & 0.335 & 0.636 & 0.244 & 0.515 & 0.416 & 0.625\\
DeepSite~\cite{jimenez2017deepsite} & $\backslash$ & 0.564 & $\backslash$ & 0.456 & $\backslash$ & $\backslash$ \\
DeepSurf~\cite{Mylonas2021-DeepSurf} & 0.386 & 0.658 & 0.289 & 0.635 & 0.510 & 0.708\\
DeepPocket~\cite{Aggarwal2022-DeepPocket} & 0.399 & 0.645 & 0.456 & \textbf{0.734} & 0.644 & 0.813\\
\midrule
EquiPocket~\cite{zhang2023equipocket} & 0.423 & 0.656 & 0.337 & 0.662 & 0.545 & 0.721\\
VN-EGNN~\cite{sestak2023vnegnn} & \textbf{0.605} & \textbf{0.750} & \textbf{0.532} & 0.659 & \textbf{0.669} & \textbf{0.820}\\
\midrule \bottomrule
\end{tabular}
\label{tab:binding site prediction benchmark}
\end{table}

\subsubsection{Limitations and Future Directions}
Although remarkable success has been achieved by applying geometric deep learning for binding site prediction, there are two limitations of existing methods that must be addressed in future research. The first is to predict the binding site conditioned on the binding ligands. Since binding ligands have various biochemical characteristics, such as varying size, polar and hydrophobic groups, binding pockets have specificity to ligands, and it is reasonable to consider ligand information in binding site prediction.
The second open question is to predict cryptic pockets that are not apparent in experimentally determined structures. However, caused by protein structural fluctuations \cite{amaro2019will, knoverek2019advanced}, ligands can bind to cryptic pockets and modulate protein function via inhibition or activation. Therefore, the ability to accurately and rapidly predict cryptic pockets is an important opportunity to expand the space of druggable proteome.  PocketMiner\cite{meller2023predicting} is a pioneering work on cryptic pocket prediction, and we expect to see more research in this area.

\begin{table*}
  \centering
\caption{Training setting, protein flexibility, and search space of current GDL molecular docking methods
}
\begin{adjustbox}{width=1\textwidth}
\renewcommand{\arraystretch}{1.1}
\label{tab:detailed molecular docking methods}
\begin{tabular}{lcccccc}
\toprule
Method & Date & Training set & Protein Representation & Side-chain Flexbility & Backbone Flexbility & Serach Space \\
\midrule
EquiBind~\cite{stark2022equibind} & Feb 2022 & PDBbind v2020 Timesplit Set (N=16,379) & Only $C_{\alpha}$ & \ding{55} & \ding{55} & Blind \\
TANKBind~\cite{lu2022tankbind} & June 2022 & PDBbind v2020 Timesplit Set (N=17,787) & Only $C_{\alpha}$ & \ding{55} & \ding{55} & Blind \\
E3Bind~\cite{zhang2022e3bind} & Oct 2022 & PDBbind v2020 Timesplit Set (N=16,379) & Only $C_{\alpha}$ & \ding{55} & \ding{55} & Blind \\
DiffDock~\cite{corso2022diffdock} & Feb 2023 & PDBbind v2020 Timesplit Set (N=17,347) & Only $C_{\alpha}$ & \ding{55} & \ding{55} & Blind \\
FABind~\cite{Pei2023FABindFA} & Oct 2023 & PDBbind v2020 Timesplit Set (N=17,299) & Only $C_{\alpha}$ & \ding{55} & \ding{55} & Blind \\
DiffDock-L~\cite{Corso2023-da-dockgen} & Jan 2024 & BindingMOAD \& PDBbind v2020 Timesplit Set (N=17,299) & Only $C_{\alpha}$ & \ding{55} & \ding{55} & Blind \\
FABind+~\cite{Gao2024FABindEM} & Mar 2024 & PDBbind v2020 Timesplit Set (N=17,299) & Only $C_{\alpha}$ & \ding{55} & \ding{55} & Blind \\
\midrule
UniMol~\cite{zhou2023unimol} & Feb 2023 & PDBbind v2020 General Set (N=16,563) & All Atom & \ding{55} & \ding{55} & 6.0 \AA~Buffer Pocket \\
EDM-Dock~\cite{masters2023deep} & Mar 2023 & BioLiP (N=47,095) & Only $C_{\alpha}$ & \ding{55} & \ding{55} & 8.0 \AA~Buffer Pocket \\
KarmaDock~\cite{Zhang2023EfficientAA} & Sep 2023 & PDBbind v2020 General Set (N=17,242) & Only $C_{\alpha}$ & \ding{55} & \ding{55} & 12.0 \AA~Buffer Pocket \\
HelixDock~\cite{Liu2023PreTrainingOL} & Oct 2023 &  PDBbind v2020 (N$\sim$19K) & All Atom & \ding{55} & \ding{55} & Pocket \\
CarsiDock~\cite{Cai2023CarsiDockAD} & Dec 2023 & PDBbind v2020 General Set (N=16,563) & All Atom & \ding{55} & \ding{55} & 5$\sim$7 \AA~Buffer Pocket \\
UniMol-V2~\cite{Alcaide2024UniMolDV} & May 2024 & BindingMOAD & All Atom & \ding{55} & \ding{55} & 10.0 \AA~Buffer Pocket \\
\midrule
DiffDock-Pocket~\cite{plainer2023diffdockpocket} & Sep 2023 & PDBbind v2020 Timesplit Set (N=16,379) & All Atom & \ding{51} & \ding{55} & 10.0 \AA~Buffer Pocket \\
DiffBindFR~\cite{Zhu2023DiffBindFRAS} & Nov 2023 & PDBbind v2020 Timesplit Set (N=16,379) & All Atom & \ding{51} & \ding{55} & 12.0 \AA~Buffer Pocket \\
FlexPose~\cite{Dong2023EquivariantFM} & Nov 2023 & APObind(N=9,221) \& PDBbind v2020 (N=18,781) & Only $C_{\alpha}$ & \ding{51} & \ding{51}  & 10.0 \AA~Buffer Pocket \\
\midrule
NeuralPLexer~\cite{qiao2022state} & Sep 2022 & PDB PL2019-74k (N=74,477) & All Atom & \ding{51} & \ding{51} & Blind \\
DynamicBind~\cite{lu2023dynamicbind} & Aug 2023 & PDBbind v2020 Timesplit Set (N=12,807) & Only $C_{\alpha}$ & \ding{51} & \ding{51} & Blind \\
Rosettafold All-Atom~\cite{krishna2023generalized} & Oct 2023 & - & All Atom & \ding{51} & \ding{51} & Blind \\
UMol~\cite{Bryant2023StructurePO} & Nov 2023 & PDBbind v2020 (N=16,420) & All Atom & \ding{51} & \ding{51} & Blind \\
AlphaFold3~\cite{Abramson2024AccurateSP} & May 2024 & - & All Atom & \ding{51} & \ding{51} & Blind \\
Chai-1~\cite{chai1} & Sep 2024 & - & All Atom & \ding{51} & \ding{51} & Blind \\
\bottomrule
\end{tabular}
\renewcommand{\arraystretch}{1}
\end{adjustbox}
\end{table*}

\subsection{Binding Pose Prediction}\label{sec:binding_pose}
\subsubsection{Problem Formulation}
Predicting the binding mode of a ligand molecule to a target protein, commonly referred to as molecular docking, is a fundamental challenge in drug discovery with a wide range of practical applications.
We can represent the target protein structure as $\mathcal{P}$, the ligand's 2D graph as $G$, and the 3D structure of the ligand as $R$. The primary goal is to develop a model for $p(R|G, \mathcal{P})$ model that can be used to predict the 3D binding pose of the ligand. 

\subsubsection{Representative Methods}
Traditional molecular docking methods rely on manually designed scoring functions and extensive conformation sampling to predict the optimal binding conformation. However, the field has undergone a transformative shift with the advent of GDL. This powerful approach leverages the inherent geometric properties of molecules and proteins, leading to significant advancements in docking accuracy and efficiency.

Besides basic information about models provided in Table.~\ref{tab:sbdd}, we further collect important and detailed characteristic of GDL docking methods in Table.~\ref{tab:detailed molecular docking methods}, including training set, protein presentation methods, search space, and whether considering protein side-chain flexibility and backbone flexibility. 

According to whether account for protein flexibility, molecular docking methods can be broadly categorized into two classes: \textbf{rigid docking} and \textbf{flexible docking}. And for whether the binding pocket is given, docking settings can be divided into \textbf{blind docking} and \textbf{site-specific docking}. Generally, the research interest transits from rigid docking to flexible docking. And in this section, we will also introduce the representative docking methods following the trend.

One noteworthy progress is the propose of rigid docking method EquiBind~\cite{stark2022equibind}, which marks the first instance of incorporating a geometric deep learning model into the task of molecular docking. 
Specifically, EquiBind~\cite{stark2022equibind} employs an SE(3)-equivariant geometric deep learning model to facilitate direct-shot predictions of both the receptor's binding site location and the bound pose of the ligand. This is achieved by predicting and aligning key pocket landmarks on the ligand and the protein. In comparative evaluations against traditional methods like VINA~\cite{Trott2010AutoDockVI} and SMINA, EquiBind substantially enhances docking efficiency, outperforming them by orders of magnitude—by a factor ranging from 3 to 5.
TANKBind~\cite{ProfileShuya2022structure} further improves over EquiBind by combining a divide-and-conquer strategy and a Trigonometry-Aware Neural network. TANKBind predicts the inter-molecular distance matrix and then takes a numerical approach to generate specific ligand coordinates based on the inter-molecular distance matrix, coordinates of protein nodes, and the pair distance matrix of ligand nodes.
The methods of TANKBind to predict complex structure by predicting inter-molecular distance matrix inspires a series further works, including UniMol~\cite{zhou2023unimol} and CarsiDock~\cite{Cai2023CarsiDockAD}.  Beyond predicting the inter-molecular distance matrix, E3Bind~\cite{zhang2022e3bind}, FABind~\cite{Pei2023FABindFA}, and FABind+~\cite{Gao2024FABindEM} directly predict the ligand coordinates iteratively. These methods designs SE(3)-equivariant graph network to predict the coordinates without the employ of a numerical approach-based generation process.
Generally, these methods treat the binding pose generation task as a regression problem.

\begin{figure}[t]
\centering
\includegraphics[width=.9\linewidth]{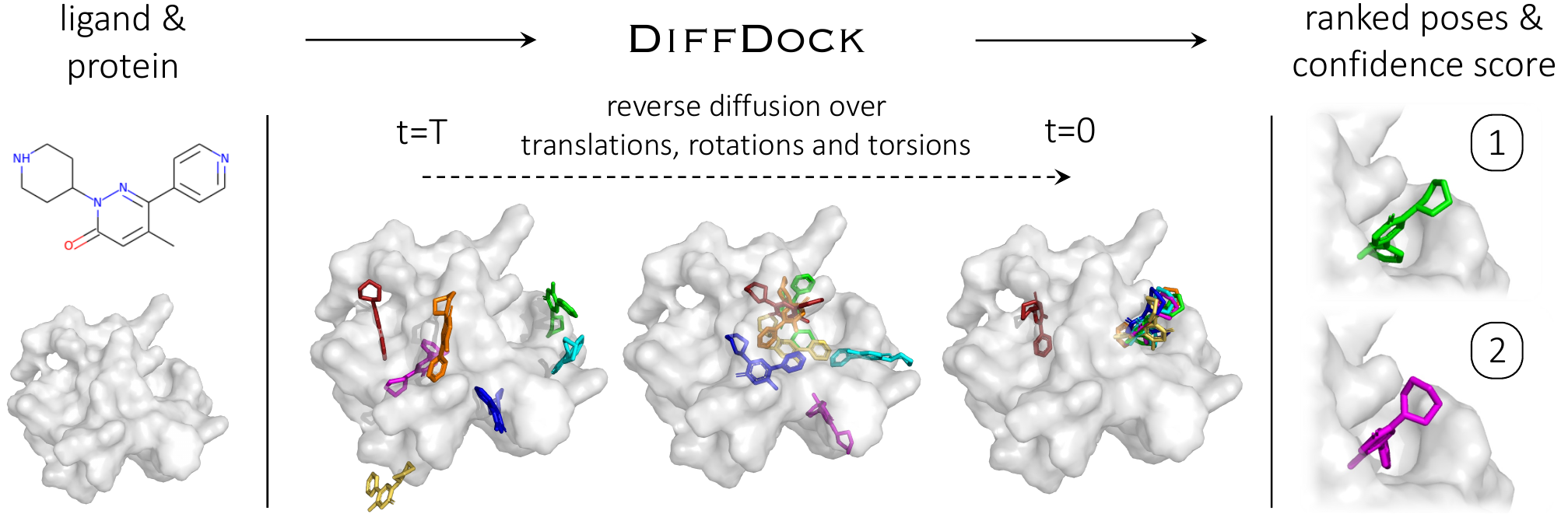}
\caption[]{
Overview of DiffDock~\cite{corso2022diffdock} for binding pose prediction. The model takes as input the separate ligand and protein structures. Randomly sampled initial poses are denoised via a reverse diffusion process over translational, rotational, and torsional degrees of freedom. A trained confidence model ranks the sampled poses to produce a final prediction and confidence score.}
\label{fig:diffdock}
\end{figure}

In contrast to the prevalent approach, DiffDock~\cite{corso2022diffdock} introduces a groundbreaking perspective by framing it as a generative modeling problem (Figure~\ref{fig:diffdock}).
As a diffusion generative model over the non-Euclidean manifold of ligand poses, DiffDock maps the manifold to the product space of the degrees of ligand freedom (translational, rotational, and torsional) involved in docking and develops an efficient diffusion process on this space. The diffusion model generates a set of candidate poses for each input protein-ligand pair, and a trained confidence model is employed to pick out the most likely pose. 
Besides the generative paradigm, another important feature of DiffDock is its inductive biases on ligand structure. Specifically, all bond lengths and angles are considered fixed during the process in DiffDock. The ligand structure are updated by changing its angels of torsional bond rather than directly changing the atomic coordinates. This inductive biases enables DiffDock to generate much more reliable ligand structure than previous methods like E3Bind and FABind. These insightful designs let DiffDock achieve substantial performance gains over previous GDL methods and motivates a series of following works.

After achieving certain success in rigid docking given holo protein structure, more recently, research efforts have increasingly gravitated towards flexible docking which explicitly incorporates protein flexibility during the docking process. This shift is crucial as protein flexibility plays a vital role in ligand binding and biological activity.
Flexible docking methods, capable of predicting accurate complex structures without relying on protein holo structures, hold immense promise for practical applications.

NeuralPLexer~\cite{qiao2022state} first takes the flexibility of protein structures into consideration, resulting in superior performance in predicting protein-ligand complex structures. Briefly, the model iteratively samples the residue-level contact maps and then determines all atom coordinates of the protein-ligand complex through a diffusion model denoising the atomic coordinates. Rather than fixing protein atomic coordinates like previous rigid docking methods, NeuralPLexer denoises coordinates of all atoms, making to perform flexible docking.
DynamicBind~\cite{lu2023dynamicbind}, similar to DiffDock, goes a step further by incorporating the degrees of freedom not only in the ligand but also within the protein backbone and side chains. This comprehensive approach allows DynamicBind to achieve enhanced accuracy in flexible docking scenarios.
In contrast, DiffDock-Pocket~\cite{plainer2023diffdockpocket} and DiffBindFR~\cite{Zhu2023DiffBindFRAS} are specifically designed for site-specific docking. During the docking process, the all-atom structure of a protein pocket is considered to model the detailed interaction between protein and ligands. Different from blind docking methods, these site-specific docking methods usually only consider the protein side-chain flexibility and assume the protein backbone flexibility fixed. 
Leveraging the known binding site information to refine the docking process, these methods achieve higher accuracy in predicting ligand poses within the defined pocket. 

While traditional methods focusing solely on protein-ligand interactions have achieved some success, recent advances in unified models like AlphaFold3~\cite{Abramson2024AccurateSP}, RosettaFold-All-Atom~\cite{krishna2023generalized}, and Chai-1~\cite{chai1} have demonstrated significantly improved performance. These models predict diverse biomolecular complexes, including protein-ligand, protein-nucleic acid, and protein-protein interactions, surpassing specialized methods in accuracy. This superior performance is attributable to two key advancements: (1) architectural innovations, such as AlphaFold3's replacement of AlphaFold2's structure module with a diffusion module operating directly on atomic coordinates, eliminating the need for rotational frame transformations or equivariant processing, leading to enhanced efficiency and accuracy; and (2) training on substantially larger and more diverse datasets compared to specialized docking models. This expansive training enables the unified models to learn more generalizable principles governing intermolecular interactions, leading to more robust and accurate predictions across a wider range of molecular complexes. The success of these unified models suggests a promising future for binding pose generation, paving the way for more efficient and effective drug discovery pipelines. 

\begin{figure}[t]
\centering
\includegraphics[width=1.0\linewidth]{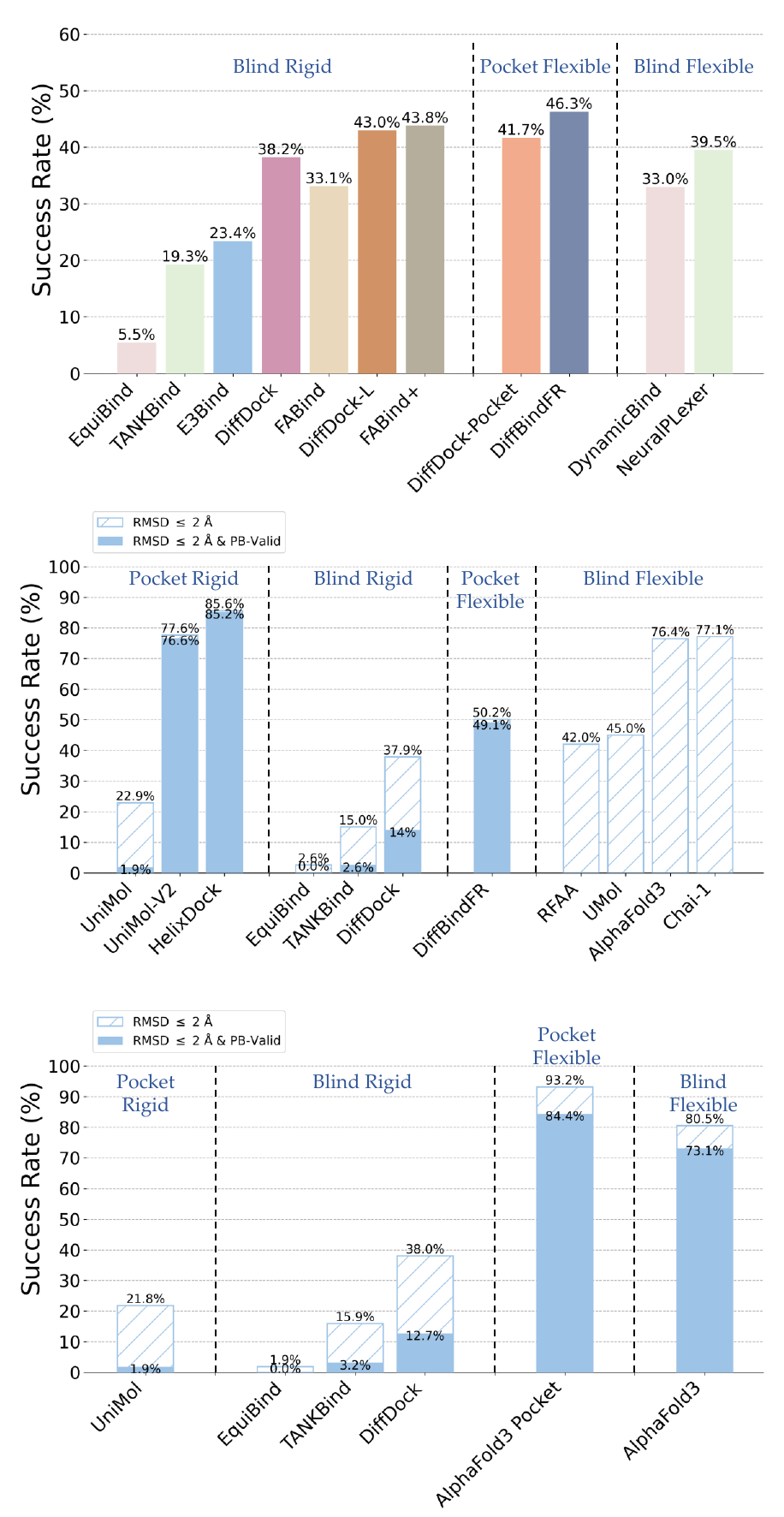}
\caption[]{Summary of docking success rate (RMSD $\leq$
2.0 \AA ) of representative molecular docking methods on PDBbind (top subfigure), PoseBusters v1 (middle subfigure), and PoseBusters v2 (bottom subfigure) datasets. Within the PoseBusters datasets, PB-Valid indicates poses that passed all chemical checks implemented in PoseBusters.}

\label{fig:docking_performance_total}
\end{figure}

\subsubsection{Datasets}
As detailed in Table.~\ref{tab:detailed molecular docking methods}, the various molecular docking methods were trained on different datasets. A comprehensive description of each dataset employed follows.

\textbf{PDBbind}~\cite{liu2017forging} is a subset of the PDB~\cite{Berman2000ThePD} that contains experimentally measured 3D structures of protein-ligand complexes. This dataset is generally used for molecular docking tasks. The newest version, PDBbind v2020, contains 19,443 protein-ligand complexes with 3,890 unique receptors and 15,193 unique ligands. Several strategies exist for partitioning this dataset into training, validation, and test sets. These include time splitting based on the release date of the structures, similarity-based splitting based on protein sequence identity, and the dataset's intrinsic general/refined/core set partitioning. Generally, different methods adopt different strategy, leading to different training and testing sets.

\textbf{PoseBusters}~\cite{Buttenschoen2023PoseBustersAD} is a dataset specialized designed for assessing the chemical and physical plausibility of ligand poses, complementing accuracy-based metrics like RMSD. The PoseBusters test suite consists of 18 checks in total, organized into three groups of tests to evaluate chemical, intramolecular, and intermolecular validity. Currently, there are two versions of PoseBusters, namely v1 and v2. Compared with v1 dataset, the v2 dataset further remove ligands which are within 5.0 \AA ~of any protein symmetry mate. These two version dataset contain 428 and 308 data respectively.

Beyond PDBbind and PoseBusters, \textbf{BioLiP}~\cite{Yang2013-gp-biolip}, \textbf{BindingMOAD}~\cite{hu2005binding}, and \textbf{APOBind}~\cite{Aggarwal2021APObindAD} provide additional protein-ligand complex data for training docking models. These datasets are often leveraged when larger training sets are required than PDBbind alone can provide.

\subsubsection{Evaluation Metrics}
\textbf{Centroid Distance} calculates the distance between the averaged coordinates of the predicted and truly bound ligand atoms. 

\noindent\textbf{Ligand Root Mean Square Deviation (L-RMSD)} is the mean squared error between the atoms of the predicted and bound ligands.
Formally, let $R\in \mathbb{R}^{n\times 3}$ and $\hat{R}\in \mathbb{R}^{n\times 3}$ be the predicted and the ground truth ligand coordinates, where $n$ is the number of atoms. The L-RMSD is obtained with:
\begin{equation}
 \text{L-RMSD}(R,\hat{R}) = \big(\frac{1}{n}\sum_{i=1}^{n}||R_i - \hat{R}_i||^2 \big)^{\frac{1}{2}},
 \label{rmsd}
\end{equation}
where $R_i$ and $\hat{R}_i$ denote the coordinate of the $i$-th atom.

\noindent\textbf{Kabsch RMSD} is the lowest possible RMSD that the roto-translation transformation of the ligand can obtain. It first uses the Kabsch algorithm to superimpose the two structures and then calculates the RMSD score similar to Equation.~\ref{rmsd}.


\subsubsection{Benchmark Performance}
Figure \ref{fig:docking_performance_total} presents a comparative performance analysis of prominent molecular docking methods.
Performance data are collected from the respective publications for each method.
Direct comparison across all methods is challenging due to variations in docking protocols and training datasets. Notably, methods such as DiffDock and DynamicBind utilize the PDBbind timesplit dataset for training and evaluation, whereas others, including CarsiDock and KarmaDock, employ the PDBbind general set. 
Despite these inconsistencies, we have endeavored to present a comprehensive and balanced comparison of the most widely-used docking methods.

The methods were evaluated using three datasets: PDBbind timesplit test set, PoseBusters v1, and PoseBusters v2, as illustrated in Figure \ref{fig:docking_performance_total}. Site-specific docking methods generally exhibit higher docking success rates compared to blind docking methods, underscoring the importance of pocket information for accurate binding pose prediction. Furthermore, current flexible docking methods, especially the unified structure prediction methods such as RosettaFold-All-Atom, AlphaFold3, and Chai-1, demonstrate significantly improved performance compared to rigid docking methods. 
Among methods specifically designed for blind docking, DiffDock exhibits the highest performance. Considering all docking settings, AlphaFold3 and Chai-1 are the top performers undoubtedly.



\subsubsection{Limitations and Future Directions}
Since the introduction of EquiBind ~\cite{stark2022equibind}, which integrated geometric deep learning into molecular docking tasks, numerous methods have emerged to address various docking scenarios, including blind, site-specific, rigid, and flexible docking. Notably, recent developments like AlphaFold 3 and Chai-1 have achieved significant progress in flexible docking, delivering highly accurate and physically plausible binding poses.  Both AlphaFold 3 and Chai-1 have been recently made open-source, enhancing accessibility for the broader scientific community. Future research should focus on creating more robust open-source docking solutions, potentially drawing inspiration from the architectures of these leading models and the progress of LLMs. Additionally, expanding docking methodologies to tackle more complex scenarios, such as multi-ligand docking and dual-target interactions, is a promising direction for further investigation.

\subsection{\emph{De Novo} Ligand Generation}\label{sec:ligand_generation}
\subsubsection{Problem Formulation}
The goal of \emph{de novo} ligand generation is to generate valid 3D molecular structures that can fit and bind to specific protein binding sites. \textit{De novo} generation involves generating a molecule while no reference ligand molecule is given, i.e., developing molecules from scratch.
Formally, let $\mathcal{P}$ denote the protein structure and $\mathcal{G}$ be the 3D ligand molecule. The objective is to learn a conditional generative model $p(\mathcal{G}|\mathcal{P})$ to capture the distribution of protein-ligand pairs.

\subsubsection{Representative Methods} 
Early methods on \emph{de novo} ligand generation represent the target protein as a 3D grid and employ 3D CNN as the encoder. For example, 
LiGAN \cite{ragoza2022generating} uses a conditional variational autoencoder trained on atomic density grid representations of protein-ligand structures for ligand generation. The molecular structures of ligands are then constructed by further atom fitting and bond inference from the generated atom densities. However, as a preliminary work, LiGAN does not satisfy the desirable equivariance property. 

The follow-up methods represent the target protein and ligand as 3D graphs/point clouds, and the equivariance is achieved by leveraging equivariant GNNs for context encoding. For example, 3DSBDD \cite{luo2021autoregressive} uses SchNet \cite{schutt2017schnet} to encode the 3D context of binding sites and estimate the probability density of atom’s occurrences in 3D space. The atoms are sampled auto-regressively until there is no room for new atoms. GraphBP \cite{liu2022generating} adopts the framework of normalizing flow \cite{rezende2015variational} and constructs local coordinate systems to predict atom types and relative positions. 

Pocket2Mol \cite{peng2022pocket2mol} adopts the geometric vector perceptrons \cite{jing2021equivariant} and the vector-based neural network \cite{deng2021vector} as the context encoder. Inspired by AlphaFold \cite{jumper2021highly} for protein structure prediction, Pocket2Mol incorporates a triangular self-attention in the encoder, where the attention bias is designed to capture the geometric constraints. Pocket2Mol jointly predicts frontier atoms, atomic positions, atom types, and covalent chemical bonds. With vector-based neurons, Pocket2Mol can efficiently sample drug molecules from tractable distributions without relying on MCMC.

By leveraging the chemical priors of molecular fragments such as functional groups, FLAG \cite{zhang2023molecule} and DrugGPS \cite{zhang2023drug} propose to generate ligand molecules fragment-by-fragment and yield more realistic substructures. For example, in FLAG \cite{zhang2023molecule}, a motif vocabulary is firstly constructed by preprocessing the dataset and extracting molecular fragments with high occurrence frequencies (i.e., motif). Drug molecules are constructed auto-regressively in the generation process with motifs as the building blocks. At each generation step, as shown in Figure~\ref{fig:FLAG}, a 3D
graph neural network encodes the intermediate context information, selects the focal motif,
predicts the next motif type, and attaches the new motif to the generated molecule. Since the bond lengths/angles are largely determined, FLAG leverages cheminformatics tools \cite{bento2020open} to effectively determine them and focus on training neural networks for rotation angle prediction.

Building based on FLAG \cite{zhang2023molecule}, DrugGPS \cite{zhang2023drug} further considers the generalizability issue of structure-based drug design models: the amount of high-quality protein-ligand complex data is rather limited and the target protein pocket may not be in the training dataset. The trained model struggles to generate good drug candidates for the unseen target protein. DrugGPS\cite{zhang2023drug} effectively incorporates the protein subpocket prior to generalizable drug molecule generation. Although two protein pockets might be dissimilar overall, they may still bind the same fragment if they share similar subpockets \cite{kalliokoski2013subpocket}. To capture the subpocket-level similarities/invariance among the
binding pockets, DrugGPS\cite{zhang2023drug} proposes to learn subpocket prototypes
and construct a global interaction graph to model the subpocket prototype-molecular motif interactions during training.

Recently, motivated by the powerful generation capability of the Diffusion models, Diffusion-based methods such as DiffSBDD~\cite{schneuing2022structure}, TargetDiff~\cite{guan2023d}, and DecompDiff~\cite{guan2023decompdiff} are proposed for non-autoregressive ligand generation and achieve superior performance. For example, TargetDiff~\cite{guan2023d} learns a joint drug molecule generative process of both continuous atom coordinates and categorical atom types with an SE(3)-equivariant network conditioned on the protein pocket. Further studies show that TargetDiff~\cite{guan2023d} can also extract representative features from protein-ligand complexes to estimate the binding affinity, providing an effective virtual screening method. Inspired by pharmaceutical practices, DecompDiff~\cite{guan2023decompdiff} considers different roles of atoms in the ligand and decomposes the ligand molecule and prior into two parts, namely arms and scaffold for drug design. The arms are responsible for the interactions with the binding regions for higher affinity, whereas the scaffold's role involves placing the arms accurately within the intended binding regions. 
Moreover, DecompDiff~\cite{guan2023decompdiff} incorporates both bond diffusion in the model and additional validity guidance in the sampling phase to improve the
properties of the generated molecules.

\begin{table*}
    \centering
    \caption{Comparison of the properties of the reference molecules and the generated molecules by different \emph{de novo} ligand generation methods. Vina Score indicates the Vina scoring function is directly calculated without ligand redocking or local optimization. Vina Min denotes that the ligand is locally minimized. Vina Dock indicates the ligand is locally optimized and redocked.}
    \begin{adjustbox}{width=1\textwidth}
    \renewcommand{\arraystretch}{1.1}
\begin{tabular}{c|cc|cc|cc|cc|cc|cc|cc}
\toprule
\diagbox{Model}{Metric} & \multicolumn{2}{c|}{Vina Score ($\downarrow$)} & \multicolumn{2}{c|}{Vina Min ($\downarrow$)} & \multicolumn{2}{c|}{Vina Dock ($\downarrow$)} & \multicolumn{2}{c|}{High Affinity ($\uparrow$)} & \multicolumn{2}{c|}{QED ($\uparrow$)}   & \multicolumn{2}{c|}{SA ($\uparrow$)} & \multicolumn{2}{c}{Diversity ($\uparrow$)} \\
 & Avg. & Med. & Avg. & Med. & Avg. & Med. & Avg. & Med. & Avg. & Med. & Avg. & Med. & Avg. & Med.  \\
\midrule
Reference   & -6.36 & -6.46 & -6.71 & -6.49 & -7.45 & -7.26 & -  & - & 0.48 & 0.47 & 0.73 & 0.74 & -    & - \\
\midrule
liGAN \cite{masuda2020generating}       & - & - & - & - & -6.33 & -6.20 & 21.1\% & 11.1\% & 0.39 & 0.39 & 0.59 & 0.57 & 0.66 & 0.67 \\
GraphBP \cite{liu2022graphbp}    & - & - & - & - & -4.80 & -4.70 & 14.2\% & 6.7\% & 0.43 & 0.45 & 0.49 & 0.48 & \textbf{0.79} & \textbf{0.78} \\
3DSBDD \cite{luo20213d}          & \textbf{-5.75} & -5.64 & -6.18 & -5.88 & -6.75 & -6.62 & 37.9\% & 31.0\% & 0.51 & 0.50 & 0.63 & 0.63 & 0.70 & 0.70 \\
Pocket2Mol \cite{peng2022pocket2mol}  & -5.14 & -4.70 & -6.42 & -5.82 & -7.15 & -6.79 & 48.4\% & 51.0\% & 0.56 & 0.57 & 0.74 & \textbf{0.75} & 0.69 & 0.71 \\
TargetDiff \cite{guan2023d}  & -5.47 & \textbf{-6.30} &-6.64 &-6.83 & -7.80&-7.91 & 58.1\% & 59.1\% & 0.48 & 0.48 & 0.58 & 0.58 & 0.72 & 0.71 \\
FLAG \cite{zhang2023molecule} &-5.30&-5.89 &-6.46 &-6.68 &-7.25 & -7.17&53.7\% & 54.8\%& 0.50 & 0.51 & \textbf{0.75} & 0.72 & 0.70 & 0.73\\
DrugGPS\cite{zhang2023drug} &-5.45&-5.81 &-6.49 &-6.88 & -7.36 & -7.42& 54.9\% & 55.7\%& \textbf{0.59} & \textbf{0.58} & 0.72 & 0.73 & 0.71 & 0.74\\
Decompdiff\cite{guan2023decompdiff} &-5.67&-6.04 &\textbf{-7.04} & \textbf{-7.09} & \textbf{-8.39}&\textbf{-8.43} & \textbf{64.4\%}& \textbf{71.0\%}& 0.45 & 0.43 & 0.61 & 0.60 & 0.68 & 0.68\\
\bottomrule
\end{tabular}
\renewcommand{\arraystretch}{1}
    \end{adjustbox}
    \label{tab:ligand generation}
\end{table*}

\begin{figure}[t]
\centering
\includegraphics[width=.98\linewidth]{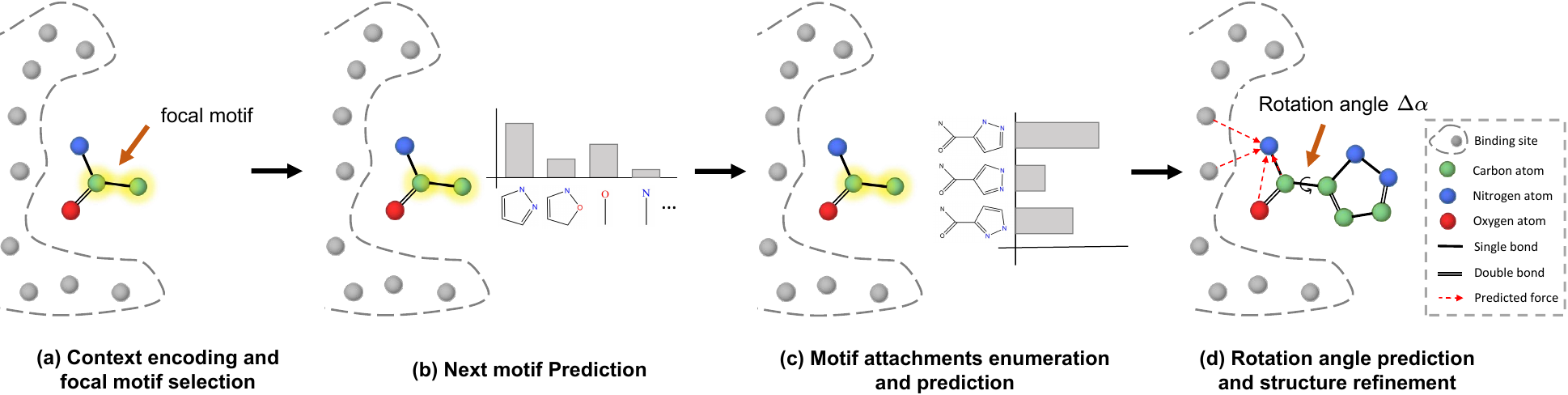} \label{fig:FLAG}
\caption[Caption for LOF]{
Overview of FLAG \cite{zhang2023molecule} for \emph{de novo} ligand generation. There are four steps in each iteration: (a) context encoding and focal motif selection, (b) next motif prediction, (c) motif attachments enumeration and prediction, and (d) rotation angle prediction and structure refinement.}
\label{fig:FLAG}
\end{figure}

\subsubsection{Datasets}
\textbf{CrossDocked} dataset \cite{francoeur2020three} is widely used in structure-based \emph{de novo} ligand design~\cite{luo20213d, peng2022pocket2mol}, which contains 22.5 million protein-molecule structures by cross-docking the Protein Data Bank \cite{berman2000protein}.  Considering the variability in the cross-docked complex qualities, existing methods typically employ filtering steps.
After filtering out data points whose
binding pose RMSD is greater than 1 {\AA}, a refined subset with around 180,000 data points is obtained. For the dataset split, 
mmseqs2 \cite{steinegger2017mmseqs2} is widely used to cluster data at 30$\%$ sequence identity, 100,000
protein-ligand pairs are randomly drawn for training and 100 proteins from the remaining clusters for testing. One hundred molecules for each protein pocket in the test set are sampled to evaluate generative models.

\noindent\textbf{Binding MOAD} \cite{hu2005binding} contains experimentally determined complexed protein-ligand structures. The dataset is filtered and split based on the proteins’ enzyme commission number \cite{bairoch1994enzyme}. Specifically,
the split ensures different sets do not contain
proteins from the same Enzyme Commission Number (EC Number) main class. Finally, there are 40,354 protein-ligand pairs
for training and 130 for testing.

\subsubsection{Evaluation Metrics}
The following metrics are widely used in related works \cite{luo20213d, peng2022pocket2mol, liu2022generating, zhang2023drug, zhang2023molecule}
to evaluate the qualities of the sampled molecules: 
(1) \textbf{Validity} is the percentage of chemically valid molecules among all generated molecules. A molecule is valid if it can be sanitized by RDkit \cite{bento2020open}.
(2) \textbf{Vina Score} measures the binding affinity between the generated molecules and the protein pockets. It can be calculated with traditional docking methods such as AutoDock Vina~\cite{trott2010autodock, alhossary2015fast} or trained CNN scoring functions \cite{ragoza2017protein}. Before calculating the vina score, the generated molecular structures are refined by universal force fields \cite{rappe1992uff}.
(3) \textbf{High Affinity} is calculated as the percentage of pockets whose generated molecules have higher affinity to the references in the test set. (3) \textbf{QED} measures how likely a molecule is a potential drug candidate. (4)  \textbf{Synthetic Accessibility (SA)} indicates the difficulty of drug synthesis (the score is normalized between 0 and 1, and higher values indicate more accessible synthesis). 
(5) \textbf{LogP}  is the octanol-water partition coefficient (LogP values should be between -0.4 and 5.6 to be promising drug candidates \cite{ghose1999knowledge}). 
(6) \textbf{Lipinski (Lip.)} calculates how many rules the molecule obeys the Lipinski’s rule
of five \cite{lipinski2012experimental}. (7) \textbf{Sim. Train} represents the Tanimoto similarity \cite{bajusz2015tanimoto} with the most similar molecules in the training set. (8) \textbf{Diversity (Div.)} measures the diversity of generated molecules for a binding pocket (It is calculated as 1 - average pairwise Tanimoto similarities). (9) \textbf{Time} records the cost of generating 100 valid molecules for a pocket. 

\subsubsection{Bechmark Performance}
We benchmark representative \emph{de novo} ligand generation methods on the CrossDocked dataset in Table~\ref{tab:ligand generation}. Generally, there is not a single method that is optimal on all the metrics. We can observe that diffusion-based methods achieve the best performance on binding affinity (Vina-related metrics). This may be attributed to their non-autoregressive generation scheme that facilitates global optimization. As for QED and SA, fragment-based methods such as DrugGPS achieve the most competitive performance. This may be explained by incorporating drug-like fragments, effectively increasing drug-likeliness and synthesizability.  

\subsubsection{Limitations and Future Directions}
Despite the success of applying geometric deep learning for \emph{de novo} ligand generation, it is still challenging to explore the vast chemical space and generate high-quality drug candidates with satisfied properties. Generally, current methods have the following limitations and require further explorations: (1) failing to consider essential chemical priors; (2) lacking ligand optimization methods; (3) noncomprehensive evaluation metrics.

Firstly, most current methods fail to consider essential chemical priors such as molecular motifs and protein-ligand interaction patterns. Therefore, the generated ligand molecules may have invalid 3D structures and limited binding affinity with the target protein. FLAG \cite{zhang2023molecule} and DrugGPS \cite{zhang2023drug} have tried to leverage chemical priors of motifs and subpockets in the model construction. In the future, we expect more methods that leverage the chemical priors for high-quality ligand generation.

Secondly, existing works fail to explicitly optimize drug properties in the generation process. In practice, it is challenging to directly generate drug candidates satisfying a series of property constraints. A common practice is sampling promising lead compounds and then conducting lead optimization. Therefore, exploring multiple-property optimization methods for \emph{de novo} ligand generation is one future direction.

Thirdly, the current evaluation metrics are noncomprehension, and most focus on 2D molecule properties such as QED and SA. A recent work, PoseCheck \cite{harris2023benchmarking}, proposes four metrics to evaluate the generated molecules' poses, including interaction profiles, steric clashes, strain energy, and redocking RMSD. Their evaluations show that the ligand molecule generated by existing methods often exhibits nonphysical features such as steric clashes, hydrogen placement issues, and high strain energy. In the future, we expect more comprehensive evaluation metrics and more advanced ligand generative models that address the shortcomings.

\begin{figure}[t]
\centering
\includegraphics[width=.98\linewidth]{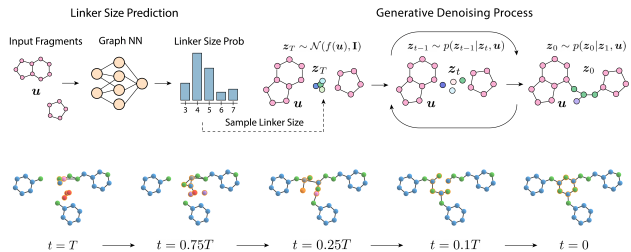}
\caption[]{Overview of DiffLinker \cite{igashov2022equivariant} for linker design. 
The inputs are the molecular fragments and the protein pocket (optional). The output is the linker that links the fragments into a complete molecule. DiffLinker is an E(3)-equivariant diffusion-based model. In the generation process, the probabilities of linker
sizes are first computed for the input fragments. Next, linker atoms are sampled and denoised using a conditioned equivariant diffusion model. The bottom shows the linker generation
process with linker atoms highlighted in orange.}
\label{fig:difflinker}
\end{figure}

\subsection{Linker Design}\label{sec:linker_design}
\subsubsection{Problem Formulation}
Designing linkers between two interested fragments is an open challenge in structure-based drug design, playing an important role in structure-based drug design.
For one hand, with a given protein pocket, computationally determining fragments that interact with the pocket is a cheaper and more efficient alternative to experimental screening methods. After identifying relevant fragments, the crucial step of linking them into a single, drug-like molecule presents considerable complexity.
For the other hand, due to the complexity of diseases, some amino acids in target proteins may mutate, leading to weak binding affinity and drug falling off.  To solve this problem, an emerging therapeutic mechanism involving proteolysis targeting chimera (PROTAC) can inhibit protein functions by prompting complete degradation of the target protein. Specifically, PROTAC is a bifunctional molecule containing two molecular fragments and a linker that links the fragments into a complete molecule. One fragment in PROTAC binds the target protein and the other fragment binds another molecule that can degrade the target protein. PROTAC only requires high selectivity in binding its targets instead of inhibiting the target protein's activity. One critical problem in PROTAC is indeed linker design.
Formally, denote the target protein as $\mathcal{P}$, the molecular fragments as $\mathcal{F}$, and the linker as $\mathcal{L}$; the objective is to learn a conditional generative model $p(\mathcal{L}|\mathcal{F, P})$.

\begin{table}[t!]
    \caption[]{Linker design performance comparisons on the ZINC test sets. Given anchors denotes that the anchors are known to the model. Sampled size indicates the linker size is sampled and is not necessarily the same as the ground truth.}
    \label{tab:linker}
    \begin{adjustbox}{width=0.5\textwidth}
    \begin{tabular}{lccccc} 
    \toprule
     Model & QED ($\uparrow$) & SA ($\uparrow$) & Valid ($\uparrow$) & Unique ($\uparrow$)& Novel ($\uparrow$) \\
    \midrule
    DeLinker\cite{imrie2020deep}              & 0.64 & 0.77  & \bf{98.3}\% & 44.2\% & \bf{47.1}\% \\
    3DLinker \cite{huang20223dlinker} (given anchors)                 & 0.65 & 0.77  & \bf{99.3}\% & 29.0\% & 41.2\% \\
    3DLinker \cite{huang20223dlinker}        & 0.65 & 0.76  & 71.5\% & 29.2\% & 41.9\% \\
    DiffLinker \cite{igashov2022equivariant} &   \bf{0.68} &   \bf{0.78}  &   93.8\% &   24.0\% &   30.3\% \\
    DiffLinker \cite{igashov2022equivariant} (given anchors)               &   \bf{0.68} &   0.77  &   97.6\% &   22.7\% &   32.4\% \\
    DiffLinker \cite{igashov2022equivariant} (sampled size)                &   0.65 &   0.76 &   90.6\% &   \bf{51.4}\% &   42.9\% \\
    DiffLinker \cite{igashov2022equivariant} (given anchors, sampled size) &   0.65 &   0.76 &   94.8\% &   \bf{50.9}\% &   \bf{47.7}\% \\
    \bottomrule
    \end{tabular}
    \end{adjustbox}
\end{table}

\subsubsection{Representative Methods}
Early work in structure-based linker design, such as DeLinker~\cite{imrie2020deep}, focusing on the sequential generation of atoms and bonds. DeLinker primarily considers simple geometric features like relative distances and orientations, ultimately producing a 2D molecular graph. A subsequent approach, 3Dlinker~\cite{huang20223dlinker} also utilizes autoregressive generation but extends this to produce 3D molecular structures. Notably, 3Dlinker achieves this without prior specification of anchor atoms (i.e., the atoms on the fragments to be linked). Instead, it sequentially predicts the anchor atom, subsequent node type, connecting bond, and 3D coordinates at each step.


In the further development of linker design methods, reinforcement learning (RL) are effectively employed.
Several RL-based approaches have emerged, including PROTAC-RL~\cite{Zheng2022AcceleratedRP}, Link-INVENT~\cite{guo2023link} and PROTAC-INVENT~\cite{chen20233d}. 
PROTAC-RL~\cite{Zheng2022AcceleratedRP} utilizes RL to train a model to generate molecules with desired pharmacokinetic~(PK) attributes, logP score, flexibility, and linker length. Link-INVENT~\cite{guo2023link} extends the widely used REINVENT model. Both PROTAC-RL and Link-INVENT, however, focus solely on 2D linker generation. In contrast, PROTAC-INVENT~\cite{chen20233d} allows for the joint sampling of 2D molecular graphs and 3D linker structures within the target protein pocket.

Diffusion models have found widespread application in linker design. Igashov et al.~\cite{igashov2022equivariant} introduced DiffLinker, a conditional diffusion model that generates molecular linkers conditioned on input fragments and, optionally, the target protein structure. As shown in Figure~\ref{fig:difflinker}, their method first computes the probability distribution of linker lengths given the input fragments, then samples and denoises linker atom positions using a conditioned equivariant diffusion model. They demonstrated that incorporating the target protein structure improves the binding affinity of the generated molecules.
Similarly, the recently proposed LinkerNet~\cite{Guan2023LinkerNetFP} employs a diffusion-based approach. However, unlike previous methods that assume fixed fragment positions, LinkerNet performs 2D-3D co-design, dynamically adjusting fragment poses during linker generation. This strategy accounts for the flexibility of protein-protein binding poses, making it particularly well-suited for applications like PROTAC design.


\subsubsection{Datasets} 
\textbf{ZINC}~\cite{sterling2015zinc} is a free database of commercially-available compounds for virtual screening. A subset of 250,000 molecules randomly selected by \cite{gomez2018automatic} is used for linker design. The dataset is preprocessed as follows: firstly, 3D conformers are generated using RDKit \cite{bento2020open}, and a reference 3D structure with the lowest energy conformation is selected for
each molecule. Then, these molecules are fragmented
by enumerating all double cuts of acyclic single bonds outside functional groups. The
results are further filtered by the number of atoms in the linker and fragments, synthetic accessibility \cite{ertl2009estimation}, ring aromaticity, and pan-assay interference compounds (PAINS) \cite{baell2010new} criteria. As a result, a single molecule may yield different combinations of two fragments separated by a linker. The ZINC dataset is randomly split into train, validation, and test sets (438,610/400/400 examples).

\noindent\textbf{CASF} \cite{su2018comparative} includes experimentally verified 3D conformations. The preprocessing procedure is the same as ZINC.

\noindent\textbf{GEOM} \cite{axelrod2022geom} is considered for real-world applications that require connecting more than two fragments with one or more linkers. The molecules are decomposed into three or more fragments with one or two linkers with an MMPA-based algorithm \cite{dossetter2013matched} and BRICS \cite{degen2008art}.

\noindent\textbf{Binding MOAD} \cite{hu2005binding} contains experimentally determined complexed protein-ligand structures. DiffLinker \cite{igashov2022equivariant} uses Binding MOAD to assess further the ability to generate valid linkers given additional information about protein pockets. 
The authors extract amino acids with at least one atom closer than 6 {\AA} to any ligand atom as the pockets. 
The molecules are preprocessed into fragments using RDKit’s implementation of MMPA-based algorithm \cite{dossetter2013matched}. The resulting dataset is split based on the proteins' Enzyme Commission (EC) numbers.

\noindent \textbf{PROTAC-DB} \cite{Ge2024PROTACDB3A} is a specialized dataset curated for training models designed to generate PROTAC-like molecules. Unlike general-purpose chemical databases such as ZINC or GEOM, which comprise diverse molecular structures, PROTAC-DB focuses exclusively on experimentally validated PROTAC molecules meticulously collected from published literature. The latest iteration, PROTAC-DB 3.0, contains 6111 PROTACs, providing a crucial resource for training and benchmarking linker design methods for PROTAC.

\subsubsection{Evaluation Metrics} 
The following metrics are widely used to evaluate linker design methods: (1) \textbf{Validity} is the percentage of chemically valid molecules among all generated molecules. (2) \textbf{Quantitative Estimation of Drug-likeness (QED)} measures how likely a molecule is a potential drug candidate. (3) \textbf{Synthetic Accessibility (SA)} indicates the difficulty of drug synthesis. (4) \textbf{Rings} is the average number of rings in the linker. (5) \textbf{Uniqueness} measures the percentage of non-duplicate generated molecules. (6) \textbf{Novelty} calculates the ratio of generated molecules not in the training set. (7) \textbf{Recovery} records the percentage of the original molecules recovered by the generation process. (8) \textbf{Root Mean Squared Deviation (RMSD)} is calculated between the generated and real linker coordinates in the cases where true molecules are recovered. (9) $\rm RD_{scit}$ \cite{putta2005conformation} evaluates the geometric and chemical similarity between the ground truth and generated molecules.

\subsubsection{Benchmark Performance}
We benchmark representative linker design methods on the ZINC dataset in Table~\ref{tab:linker}. In the default setting, all the methods generate the linker the same size as the ground truth. 
DiffLinker \cite{igashov2022equivariant} as the state-of-the-art method achieves the best results on molecule drug-likeliness (QED) and synthesizability (SA). 
We also note that sampling the linker size instead of the predefined linker size can significantly improve the novelty and uniqueness of the generated linkers without much degradation of the other important metrics.

\subsubsection{Limitations and Future Directions}

Despite significant advancements in linker design through geometric deep learning, several challenges persist. Notably, many existing methods overlook the protein pocket context, leading to generated molecules that may exhibit steric clashes or suboptimal binding affinities with target proteins. Recent approaches, such as DiffLinker, have begun to address this issue by conditioning linker design on the pocket environment.  However, a major obstacle remains: the limited availability of ternary 3D binding structures of PROTAC molecules with their two target proteins. These structures are crucial for accurately modeling the protein pocket context, yet publicly accessible datasets, like the Protein Data Bank (PDB), contain fewer than 100 such structures.  Efficient and precise computational prediction of these ternary structures continues to be an active area of research. Additionally, current 3D linker design models often assume known relative positions of the fragments, which may not be practical in real-world scenarios. Generative models like LinkerNet, which co-design both fragment poses and linkers, offer a more favorable approach in such contexts.  

\begin{table*}
  \centering
\caption{Training setting, testing setting, and test performance of binding affinity prediction models.
}
\begin{adjustbox}{width=1\textwidth}
\renewcommand{\arraystretch}{1.1}
\label{tab:affinity}
\begin{tabular}{lccccc}
\toprule
Method & Year & Training set & Testing set & RMSE ($\downarrow$) & PCC ($\uparrow$)\\
\midrule
Pafnucy~\cite{StepniewskaDziubinska2018DevelopmentAE}
& 2018 & PDBbind v2016 general set (N=11,906) & PDBbind v2016 core set (N=290) & 1.420 & 0.780 \\
DeepAtom~\cite{Li2019DeepAtomAF}
& 2019 & PDBbind v2016 refined set (N=3,390) & PDBbind v2016 core set (N=290) & 1.318 & 0.807 \\
OnionNet~\cite{Zheng2019OnionNetAM}
& 2019 & PDBbind v2016 general set (N=11,906) & PDBbind v2016 core set (N=290) & 1.287 & 0.816 \\
PIGNet~\cite{moon2022pignet} 
& 2021 & PDBbind v2019 refined augment set (N=1,656,600) & PDBbind v2016 core set (N=283) & - & 0.749 \\
Fusion~\cite{Jones2020ImprovedPB}
& 2021 & PDBbind v2016 general set (N=-)& PDBbind v2016 core set (N=290) & 1.270 & 0.820  \\
OnionNet-2~\cite{Wang2021OnionNet2AC}
& 2021 & PDBbind v2019 general set (N=17,367) & PDBbind v2016 core set (N=285) & 1.164 & 0.864 \\
IGN~\cite{jiang2021interactiongraphnet}
& 2021 & PDBbind v2016 general set (N=8,298) & PDBbind v2016 core set (N=262) & 1.220 & 0.837 \\ 
SIGN~\cite{li2021structure}
& 2021 & PDBbind v2016 refined set (N=3,390) & PDBbind v2016 core set (N=290) & 1.316 & 0.797 \\
PLIG~\cite{moesser2022protein}
& 2022 & PDBbind v2020 general set + PDBbind v2016 refined set (N=19,451) & PDBbind v2016 core set (N=285) & 1.210 & 0.840 \\
PaxNet~\cite{zhang2022efficient}
& 2022 & PDBbind v2016 refined set (N=3,390) & PDBbind v2016 core set (N=290) & 1.263 & 0.815 \\
GLI~\cite{Zhang2022PredictingPB}
& 2022 & PDBbind v2016 refined set (N=3,390) & PDBbind v2016 core set (N=290) & 1.294 &  - \\
GIGN~\cite{yang2023geometric}
& 2023 & PDBbind v2016 general set (N=11,906) & PDBbind v2016 core set (N=290) & 1.190 & 0.840 \\
KIDA~\cite{lu2023improving}
& 2023 & PDBbind v2016 general set (N=12,500) & PDBbind v2016 core set (N=285) & 1.291 & 0.837 \\
GraphscoreDTA~\cite{Wang2023GraphscoreDTAOG}
& 2023 & PDBbind v2019 general set (N=9,869) & PDBbind v2016 core set (N=279) & 1.249 &  0.831 \\
PLANET~\cite{Zhang2023PLANETAM}
& 2023 & PDBbind v2020 general set (N=15,616) & PDBbind v2016 core set (N=285) & 1.247 & 0.824 \\
MBP~\cite{yan2023multi}
& 2023 & PDBbind v2016 refined set (N=3,390) & PDBbind v2016 core set (N=290) & 1.263 & 0.825  \\ 
GET~\cite{Kong2023GeneralistET}
& 2023 & PDBbind v2019 refined set (N=3,507) & PDBbind v2019 refined set (N=490) & 1.309 & 0.633 \\ 
BindNet~\cite{Feng2023ProteinligandBR}
& 2023 & PDBbind v2019 refined set (N=3,507) & PDBbind v2019 refined set (N=490) & 1.340 & 0.634 \\ 
GIANT~\cite{Li2024GIaNtPB}
& 2024 & PDBbind v2016 refined set (N=3,390) & PDBbind v2016 core set (N=290) & 1.269 & 0.814\\ 
\bottomrule
\end{tabular}
\renewcommand{\arraystretch}{1}
\end{adjustbox}
\end{table*}

\subsection{Protein Pocket Generation}\label{sec:pocket_generation}
\subsubsection{Problem Formulation}

Protein pocket generation aims to design the protein pocket regions that interact with ligand molecules. These pockets are crucial for the protein's function and are characterized by their unique sequences and three-dimensional conformations. The task involves generating both the amino acid sequences and the corresponding 3D structures of these pockets, ensuring they are compatible with a given ligand and the surrounding protein scaffold.
Formally, let $\mathcal{P}$ represent the entire protein structure, $\mathcal{R}$ denote the region of the protein excluding the pocket, and $\mathcal{L}$ be the ligand molecule. The objective is to learn a conditional generative model $p(\mathcal{G}|\mathcal{L}, \mathcal{R})$, where $\mathcal{G}$ encompasses the sequence and structure of the protein pocket. This model captures the distribution of protein pockets conditioned on both the ligand and the existing protein scaffold. 

\subsubsection{Representative Methods}
\begin{figure}[t]
\centering
\includegraphics[width=.95\linewidth]{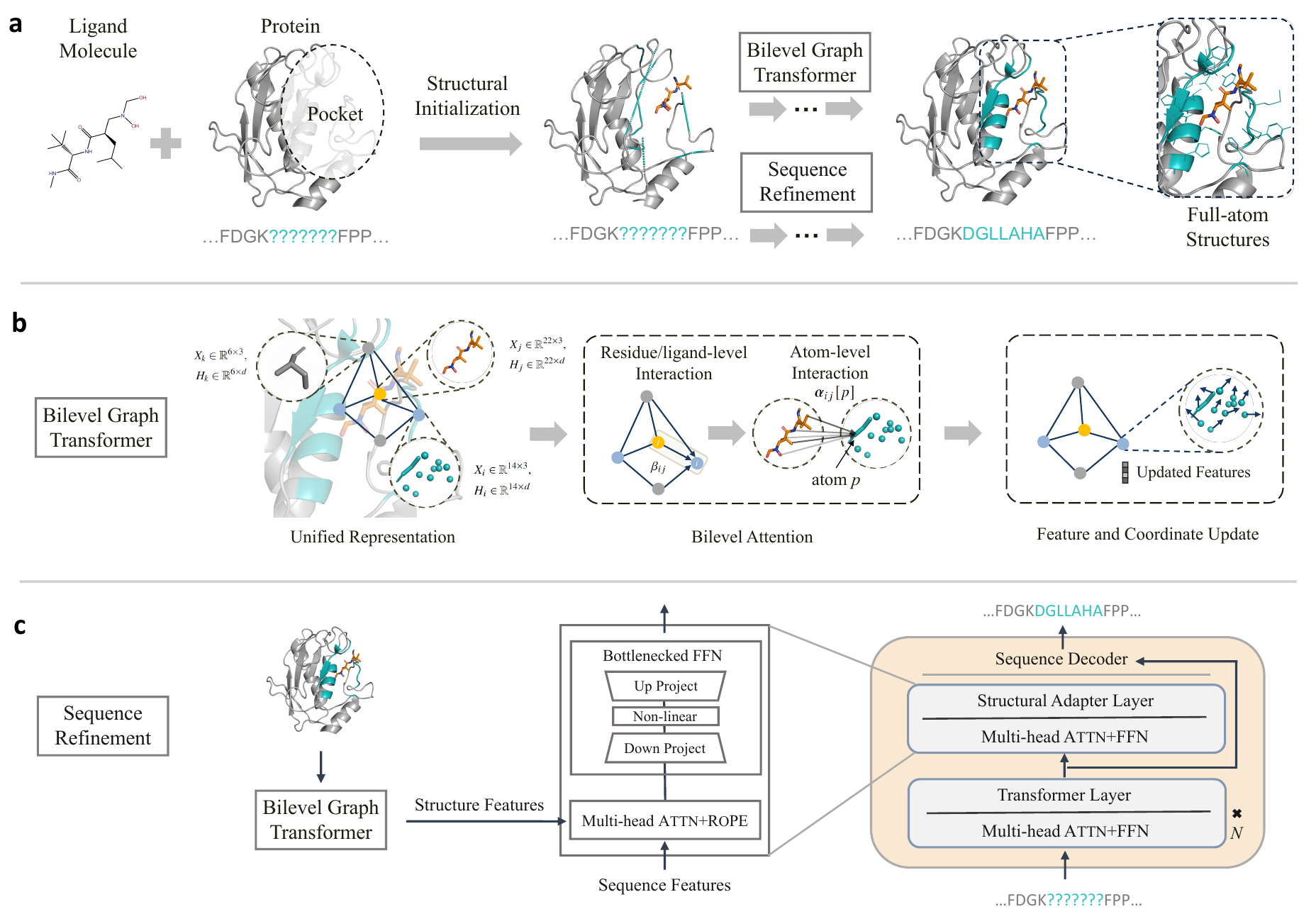}
\caption[]{
Overview of PocketGen \cite{zhang2024efficient} for protein pocket generation. A Bilevel graph transformer is leveraged in PocketGen for all-atom structural encoding and update. The sequence refinement module adds lightweight structural adapter layers into pLMs for sequence prediction.}
\label{fig:pocketgen}
\end{figure}
Recent advancements in protein pocket generation have introduced several notable methods, each employing distinct strategies to address the complexities of designing ligand-binding sites.

FAIR (Full-Atom Iterative Refinement) \cite{zhang2023full} adopts a coarse-to-fine approach, beginning with the generation of protein backbones and progressively refining to full-atom structures. This method utilizes a hierarchical context encoder alongside structure refinement modules to effectively capture inter-residue and pocket-ligand interactions, resulting in significant improvements in amino acid recovery rates and root-mean-square deviation metrics. 

Building upon the principles of sequence-structure consistency, PocketGen \cite{zhang2024efficient} employs a bilevel graph transformer to capture interactions at multiple granularities, including atom-level and residue/ligand-level. It integrates a structural adapter into a protein Language Model (pLM) for sequence refinement, enabling the generation of full-atom ligand-binding protein pockets with high fidelity. Notably, PocketGen has demonstrated superior performance in protein pocket generation tasks, achieving an average amino acid recovery rate of 63.40\% and a Vina score of -9.655 for top-1 generated protein pockets on the CrossDocked dataset. 

RFDiffusion \cite{watson2023novo} introduces a diffusion probabilistic model to generate protein backbones, focusing on designing novel protein structures by modeling the distribution of protein backbones and sampling from this distribution to create new structures. This approach has been applied to various protein design tasks, including the generation of protein backbones that can accommodate specific functional sites.
Extending the capabilities of RFDiffusion, RFDiffusionAA \cite{krishna2024generalized} incorporates side-chain modeling to facilitate full-atom protein design. By employing a diffusion model to generate both backbone and side-chain conformations, RFDiffusionAA enables the design of complete protein structures with desired properties, demonstrating improvements in generating protein structures with higher accuracy in side-chain placements compared to methods focusing solely on backbone generation.

Addressing the challenges inherent in protein pocket generation, PocketFlow \cite{zhang2024generalized} introduces a generative model that incorporates protein-ligand interaction priors based on flow matching. During training, PocketFlow learns to model key types of protein-ligand interactions, such as hydrogen bonds, and utilizes multi-granularity guidance during sampling to generate high-affinity and structurally valid pockets. Extensive experiments have shown that PocketFlow outperforms baselines on multiple benchmarks, achieving an average improvement of 1.29 in Vina Score and 0.05 in scRMSD. 

Collectively, these methods represent significant strides in the field of protein pocket generation, each introducing unique approaches to address the complexities of designing ligand-binding sites. 

\subsubsection{Datasets}
In protein pocket generation \cite{zhang2023full, zhang2024generalized}, two primary datasets are utilized:

\textbf{CrossDocked Dataset} comprises approximately 22.5 million protein-ligand complexes generated through cross-docking procedures applied to the Protein Data Bank (PDB). Due to variability in the quality of these complexes, a common practice is to filter out data points with a binding pose root-mean-square deviation (RMSD) exceeding 1 Å, resulting in a refined subset of around 180,000 data points. For dataset partitioning, tools like MMseqs2 are employed to cluster data at 30\% sequence identity. Typically, 100,000 protein-ligand pairs are randomly selected for training, while 100 proteins from distinct clusters are reserved for testing. During evaluation, 100 molecules are sampled for each protein pocket in the test set to assess generative models. 

\textbf{Binding MOAD} (Mother of All Databases) contains experimentally determined protein-ligand complex structures. To ensure diversity, the dataset is filtered and split based on the proteins' Enzyme Commission (EC) numbers, ensuring that different subsets do not contain proteins from the same EC main class. This process yields 40,354 protein-ligand pairs for training and 130 pairs for testing. 

\subsubsection{Evaluation Metrics}
In evaluating the quality of generated protein pockets, several metrics are commonly employed: (1)
\textbf{Amino Acid Recovery Rate (AAR)} assesses the percentage of amino acids in the generated pocket that match the native pocket's sequence, indicating the accuracy of sequence generation. 
(2) 
\textbf{Root-Mean-Square Deviation (RMSD)} measures the average distance between atoms in the generated pocket and those in the native structure, reflecting the structural fidelity of the generated pocket. 
(3)
\textbf{Vina Score} is calculated with AutoDock Vina, to estimate the binding affinity between the generated pocket and a target ligand, with lower scores indicating stronger predicted binding.
(4) \textbf{Binding Affinity Metrics} like \textbf{MM-GBSA} \cite{yang2023uni} and \textbf{GlideSP} scores \cite{friesner2004glide} are used to further evaluate the binding affinity between the generated pocket and the ligand, providing a comprehensive view of interaction strength.
(5) \textbf{Structural Validity} of the generated pockets using \textbf{scRMSD}, \textbf{scTM}, and \textbf{pLDDT}. The amino acid sequence for the protein pocket structure is derived using ProteinMPNN \cite{dauparas2022robust}, and the pocket structure is predicted using ESMFold \cite{lin2023evolutionary} or AlphaFold2 \cite{jumper2021highly}. The \textbf{scRMSD} is calculated as the self-consistency root mean squared deviation between the generated structure’s backbone atoms and the predicted structure. \textbf{scTM}, the self-consistency template modeling score, is calculated by comparing the TM-score \cite{zhang2004scoring} between the predicted and generated structures. Scores range from 0 to 1, with higher values indicating greater designability. \textbf{pLDDT} score indicates the confidence in structural predictions, ranging from 0 to 100. 
(6) \textbf{Computational Efficiency} records the time and resources required to generate protein pockets, highlighting the method's practicality for large-scale applications.

\subsubsection{Benchmark Performance}
Table \ref{tab:pocket} reports the statistics of the top generated protein pockets on the CrossDocked datasets. A traditional method based on physical energy optimization (i.e., PocketOpt \cite{noske2023pocketoptimizer}) is included here for reference. In the table affinity metrics such as Vina score, MM-GBSA, and GlideSP are reported. Structural validity metrics including pLDDT and scRMSD are also reported. The success rate measures the proportion of proteins for which the model generates pockets exhibiting higher binding affinity than the reference pockets in the dataset. Among the selected methods, we observe PocketGen achieves the best performance in success rates, binding affinity, and structural validity, owing to its unique design of structural encoder and pLM-enhanced sequence refinement.

\begin{table}[t!]
    \caption[]{The table presents the performance of the top 1/3/5/10 generated designable protein pockets, ranked by their Vina scores, on the CrossDocked dataset. The success rate indicates the percentage of proteins for which the model generates pockets with higher binding affinity compared to the reference pockets in the dataset. }
    \label{tab:pocket}
    \begin{adjustbox}{width=0.5\textwidth}
\begin{tabular}{l|ccccc}
\toprule
 & PocketOpt \cite{noske2023pocketoptimizer} & FAIR \cite{zhang2023full} & RFDiffusion \cite{watson2023novo} & RFAA \cite{krishna2024generalized} & PocketGen \cite{zhang2024efficient} \\
\midrule
\multicolumn{6}{c}{Top-1 generated protein pocket} \\ 
 Vina score ($\downarrow$) & -9.216 & -8.792 & -9.037 & -9.216 & {\bf -9.655} \\
MM-GBSA ($\downarrow$) & -58.754 & -51.923 & -54.817 & -59.255 & {\bf -63.542} \\
 GlideSP ($\downarrow$) & -8.612 & -7.584 & -8.485 & -8.540 & {\bf -8.916} \\
Success Rate ($\uparrow$) & 0.923 & 0.796 & 0.891 & 0.930 & {\bf 0.974} \\
 pLDDT  ($\uparrow$) & - & 83.285 & 84.432 & 86.571 & {\bf 86.830} \\
 scRMSD  ($\downarrow$) & - & 0.693 & 0.675 & 0.654 & {\bf 0.645} \\
\midrule
\multicolumn{6}{c}{Top-3 generated protein pockets} \\ 
 Vina score ($\downarrow$) & -8.878 & -8.321 & -8.876 & -8.980 & {\bf -9.353} \\
 MM-GBSA ($\downarrow$) & -53.372 & -46.050 & -52.423 & -53.593 & {\bf -60.770} \\
 GlideSP ($\downarrow$) & -8.360 & -7.348 & -8.219 & -8.233 & {\bf -8.670} \\
pLDDT  ($\uparrow$) & - & 83.025 & 84.260 & {\bf 86.289} & 86.280 \\
scRMSD  ($\downarrow$) & - & 0.692 & 0.685 & {\bf 0.659} & 0.660 \\
\midrule
 \multicolumn{6}{c}{Top-5 generated protein pockets} \\ 
 Vina score ($\downarrow$) & -8.702 & -7.943 & -8.510 & -8.689 & {\bf -9.239} \\
MM-GBSA ($\downarrow$) & -52.080 & -37.816 & -46.847 & -51.651 & {\bf -58.083} \\
 GlideSP ($\downarrow$) & -8.173 & -7.289 & -8.022 & -8.093 & {\bf -8.417} \\
 pLDDT  ($\uparrow$) & - & 83.748 & 84.505 & 85.617 & {\bf 85.969} \\
scRMSD  ($\downarrow$) & - & 0.698 & 0.680 & 0.657 & {\bf 0.655} \\
\midrule
\multicolumn{6}{c}{Top-10 generated protein pockets} \\
 Vina score ($\downarrow$) & -8.556 & -7.785 & -8.352 & -8.524 & {\bf -9.065} \\
 MM-GBSA ($\downarrow$) & -49.257 & -33.670 & -45.726 & -47.325 & {\bf -54.800} \\
 GlideSP ($\downarrow$) & -7.935 & -7.131 & -7.806 & -7.840 & {\bf -8.196} \\
 pLDDT  ($\uparrow$) & - & 83.271 & 84.080 & 85.442 & {\bf 85.945} \\
 scRMSD ($\downarrow$) & - & 0.706 & 0.688 & 0.680 & {\bf 0.659} \\
\bottomrule
\end{tabular}
    \end{adjustbox}
\end{table}

\subsubsection{Limitations and Future Directions}
Despite significant advancements in protein pocket generation, several challenges persist. Ensuring sequence-structure consistency remains complex, as accurately aligning generated amino acid sequences with their corresponding 3D structures is difficult.  Additionally, modeling intricate protein-ligand interactions, especially with diverse ligands like nucleic acids and peptides, poses substantial challenges.  Many current methods are tailored to specific ligand categories, limiting their applicability across various ligand types.  Furthermore, the resource-intensive nature of some generative models can hinder their scalability and practical application.  Addressing these challenges requires developing models that more effectively integrate sequence and structural data, creating methods capable of accurately representing a wide range of protein-ligand interactions, designing models with the flexibility to handle various ligand types, and streamlining algorithms to reduce computational demands, facilitating broader adoption in research and industry.

\subsection{Binding Affinity Prediction}\label{sec:binding_affinity}

\subsubsection{Problem Formulation}
Protein-ligand binding affinity is a measurement of interaction strength. Accurate affinity prediction helps design effective drug molecules and plays a vital role in SBDD.
Formally, denoting the bound protein structure as $\mathcal{P}$, the bound ligand as $\mathcal{L}$, and the binding affinity as $y$, our target is to train a model $f(\mathcal{P},\mathcal{L}) = y $ for binding affinity prediction.

\subsubsection{Representative Methods}
The exploration of binding affinity prediction methods has a long-standing history. Early studies focused on utilizing empirical formulas \cite{Guedes2018EmpiricalSF} or designing handcrafted features coupled with traditional machine learning algorithms for binding affinity prediction \cite{BitencourtFerreira2020MachineLS}.
Despite some advancements, these methods have limited prediction accuracy and require considerable feature engineering to perform well.

\begin{figure}[t]
\centering
\includegraphics[width=.9\linewidth]{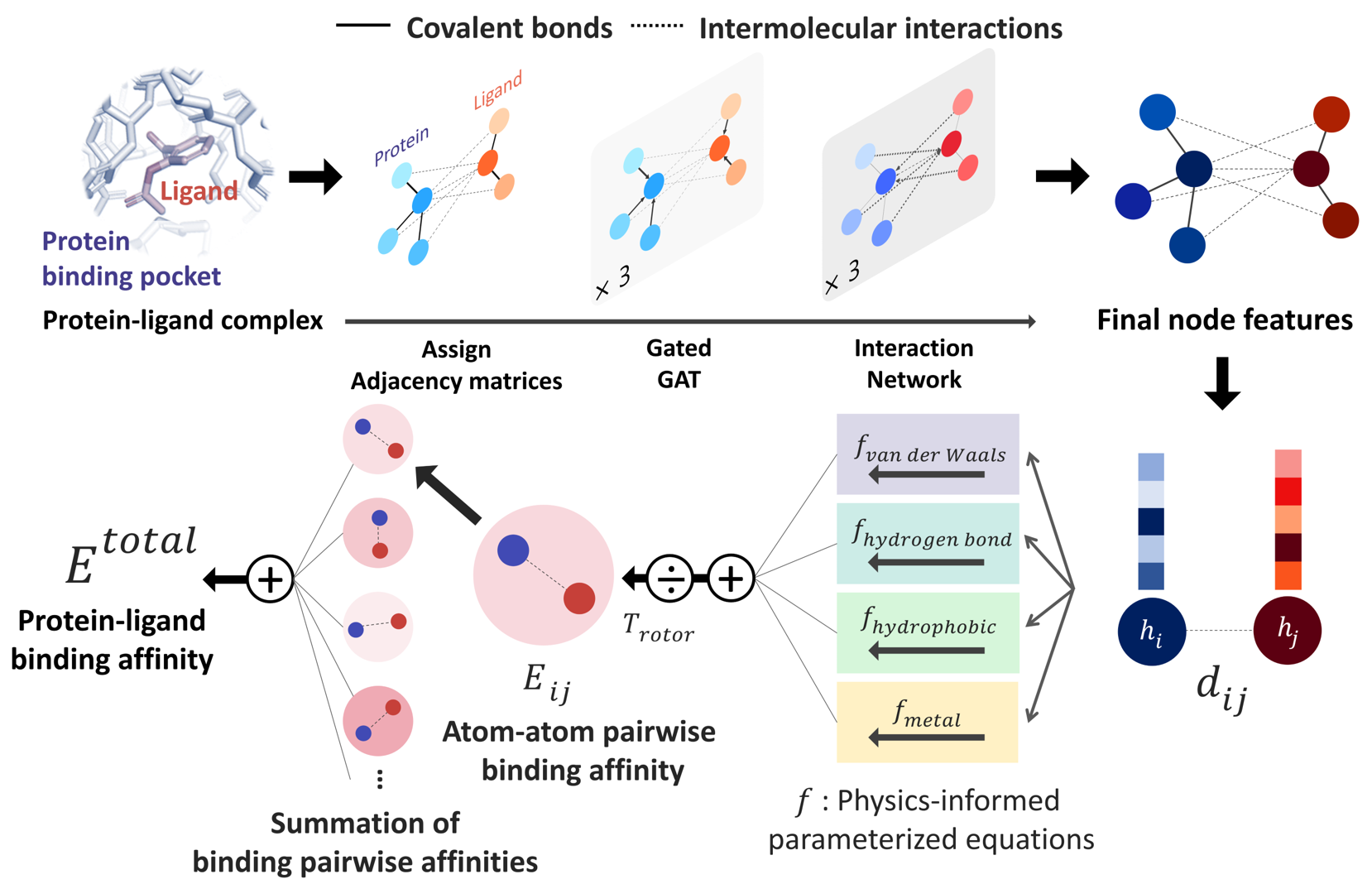}
\caption[]{
Overview of PIGNet~\cite{moon2022pignet} for binding affinity prediction. A protein-ligand complex is represented in a graph, and adjacency matrices are assigned from the binding structure of the complex. Each node feature is updated through neural networks to carry the information of covalent bonds and intermolecular interactions. Given each atom pair's distance and final node features, four energy components are calculated from the physics-informed parameterized equations. The total binding affinity is obtained as a sum of pairwise binding affinities, a sum of the four energy components divided by an entropy term.}
\label{fig:pignet}
\end{figure}

Recent research has highlighted the application of geometric deep learning methods, representing the protein-ligand complex structure as 3D grids or 3D graphs for processing and prediction.
These approaches directly model the relationship between the complex's 3D structure and binding affinity using CNNs or GNNs. 
For example, given a complex structure, 
Pafnucy~\cite{StepniewskaDziubinska2018DevelopmentAE} extracts a 20 Å cubic box focused on the geometric center of the ligand and discretizes it into a $21 \times 21 \times 21 \times 19$ grid with 1 Å resolution. A 3D-CNN is then employed to process the grid, treating it as a multi-channel 3D image.
SIGN~\cite{li2021structure} converts the complex structure into a complex 3D graph and designs a structure-aware interactive graph neural network to capture 3D spatial information and global long-range interactions using polar-inspired graph attention layers in a semi-supervised manner.
PIGNet~\cite{moon2022pignet} introduces a novel physics-informed graph neural network, which can predict accurate binding affinity based on four physics energy components – van der Waals (vdW) interaction, hydrogen bond, metal-ligand interaction, and hydrophobic interaction (Figure~\ref{fig:pignet}).
Fusion~\cite{Jones2020ImprovedPB} simultaneously utilizes the complex 3D grid representation and 3D graph to capture different characteristics of interactions.
HOLOPROT~\cite{somnath2021multi} considers both complex structures and complex surfaces. 

Besides specialized methods trained on size-limited downstream datasets for affinity prediction, recent advances in representation learning demonstrate the effectiveness of general-purpose molecular representations and pre-training strategies.
Regarding model architecture, Kong et al. propose GET~\cite{Kong2023GeneralistET}, a universal model representing arbitrary 3D complexes as geometric graphs of sets, effectively capturing robust molecular representations from both domain-specific hierarchies and domain-agnostic interaction physics. Concerning pre-training strategies, UniMol~\cite{zhou2023unimol} and BindNet~\cite{Feng2023ProteinligandBR} employ self-supervised learning, such as structure prediction and masked atom reconstruction, to exploit inherent data structure. MBP~\cite{yan2023multi} introduces the first supervised affinity pre-training framework, which involves training the model to predict the ranking of samples from the same bioassay to overcome the label noisy cased by different bioassay experimental settings. This pre-training uses a self-constructed ChEMBL-Dock dataset containing over 300,000 experimental affinity labels and about 2.8M docking-generated complex structures.
These universal molecular representation methods and pre-training strategies generally exhibit significant performance improvements over prior approaches, demonstrating their effectiveness.


\subsubsection{Datasets}
\textbf{PDBbind} \cite{liu2017forging} is the most commonly used dataset for binding affinity prediction. As previously mentioned, the latest version of the dataset consists of 19,443 complexes. Specifically, the dataset comprises three overlapping subsets: the general set (14,127 3D protein-ligand complexes), the refined set (5,316 complexes selected from the general set with higher quality), and the core set (290 complexes selected as the highest quality benchmark for testing). It is customary to train and validate models on either the general or refined sets and evaluate them on the core set. The PDBbind v2016 core set is also known as the CASF-2016 dataset.
Recently, a sequence identity-based splitting strategy~\cite{townshend2021atom3d} has also been applied. This strategy splits protein-ligand complexes such that no protein in the test dataset has more than 30\% sequence identity with any protein in the training dataset.

\noindent\textbf{CSAR-HiQ} \cite{Dunbar2013CSARDS} is another commonly used dataset, consisting of two subsets containing 176 and 167 complexes, respectively. This dataset is often used as the independent dataset in generalizability benchmarks.


\subsubsection{Evaluation Metrics}
\textbf{Root Mean Square Error (RMSE)} and \textbf{Mean Absolute Error (MAE)} are widely used to quantify the errors between the predicted values and the ground-truth values \cite{durrant2013autogrow}. These two metrics are the most direct evaluation metrics for prediction errors.

\noindent\textbf{Pearson’s Correlation Coefficient (PCC)}~\cite{yeager2019spss} quantifies the linear correlation between the predicted values and the ground-truth values. This metric serves as a means to evaluate prediction accuracy. Unlike RMSE and MAE, PCC is a normalized value ranging from -1 to 1, allowing for a standardized assessment of prediction accuracy.

\noindent\textbf{Spearman’s Correlation Coefficient (SCC)}~\cite{sedgwick2014spearman} quantifies the ranking correlation between predicted values and experimental values. It is calculated as the PCC between the rank values of the two variables. This metric is relevant, as affinity prediction is frequently employed to identify molecules with the highest rankings in virtual screening.


\subsubsection{Benchmark Performance}
The field of affinity prediction methods has a rich history, characterized by various methodologies and the utilization of diverse datasets. Consequently, conducting a fair and comprehensive comparison of all these methods presents a formidable challenge.
In Table~\ref{tab:affinity}, we present statistics on training and testing settings and the performance of these binding affinity prediction models.
As shown in the table, although the testing settings of these methods are generally the same, their training settings vary significantly. Larger training datasets generally lead to better results.
For future research in binding affinity prediction, it is essential to carefully select suitable settings for training and testing.

\subsubsection{Limitations and Future Directions}
While recent advancements in geometric deep learning for affinity prediction have significantly improved accuracy, several critical limitations remain to be addressed.
Firstly, the current geometric deep learning methods are predominantly trained on co-crystal complex structures where all the data are positive, making it challenging for these methods to effectively screen out true active ligands from a large pool of decoys. 
As exemplified by PIGNet, previous research has aimed to enhance both the scoring power and the screening power of affinity prediction models.
Nevertheless, their results suggest that improving screening power often results in a trade-off with reduced scoring power, underscoring the difficulty of simultaneously achieving high performance in both aspects.
It is essential to develop robust models capable of excelling in scoring and screening, expanding their practical applications.

Furthermore, generalization to protein structures beyond those encountered during training is crucial.
One potential solution to this challenge involves creating additional high-quality training datasets.
Currently, there are approximately 5,000 high-quality protein-ligand complexes with experimentally verified affinities, as the PDBbind refined set exemplifies.
However, this limited dataset often constrains the full training of deep learning models.
Additionally, integrating prior physical knowledge into deep learning models, such as physics-informed deep learning~\cite{Wahlstrm2021PhysicsInformedML}, represents a promising avenue for enhancing generalization capabilities.

\section{Challenges and Opportunities}
\label{sec:future}

We discuss challenges and opportunities in SBDD across various dimensions, including algorithmic innovation, practical considerations concerning model and output evaluation, and integration with experimental systems.

\subsection{Challenges}

\begin{itemize}
\item \textbf{Oversimplified Problem Formulation}: In structure-based drug design (SBDD), problem formulations must align with real-world applications and adhere to established physical and chemical principles. For instance, numerous studies on binding pose generation and \emph{de novo} ligand generation operate under the assumption that the target protein structure remains static. In reality, protein structures exhibit flexibility and can experience intrinsic or induced conformational alterations \cite{koshland1995key}. This discrepancy underscores a gap between SBDD models and their practical applications.
\item \textbf{Out-of-distribution Generalization}: Many existing studies fall short in effectively addressing the challenges of out-of-distribution (OOD) generalization. Limitations in dataset size and suboptimal dataset-splitting strategies often lead to overestimation of model performance. For example, during the COVID-19 pandemic, generative models were tasked with designing ligand molecules for novel protein targets, such as the main protease of SARS-CoV-2. These scenarios underscore the importance of developing robust and generalizable geometric deep learning models that can reliably handle real-world applications and adapt to unseen data distributions.
\item \textbf{The Need for Reliable Evaluation Metrics}: Establishing robust criteria to define an optimal drug candidate remains challenging. Even though various evaluation metrics have been proposed, they often fall short in their applicability. Some models exploit shortcuts \cite{geirhos2020shortcut} in these metrics (e.g., larger generated ligand molecule may lead to lower vina score), resulting in the generation of molecules with limited real-world utility.
\item \textbf{Lack of Large-scale Benchmarks}: While datasets and evaluation splits are available for diverse SBDD tasks, there remains a dearth of large-scale, reliable benchmarks with high-quality data. For example, the refined dataset from PDBbind \cite{liu2017forging} used in training affinity prediction models encompasses merely 5,000 complexes. The CrossDocked dataset \cite{francoeur2020three}, used to benchmark \emph{de novo} ligand design methods, comprises only 2,922 distinct proteins and 13,780 unique ligand molecules. These datasets pale compared to the size of chemical space and protein universe, underscoring the need for expansive, high-quality benchmarks.
\item \textbf{The Need for Experimental Verification}: Computationally evaluating generated drug candidates using a set of metrics, while valuable, is not sufficient. Experimental verification using \emph{in vivo} or \emph{in vitro} tests is crucial to validate a candidate's effectiveness. These experimental outcomes can be harnessed to refine the models, facilitating an integrative loop between computational simulations and empirical experiments.
\item \textbf{Lack of Interpretability}: Achieving interpretability is a paramount yet formidable task for deep learning models, often perceived as black boxes. Within SBDD, researchers often seek insights into rationales behind predicted protein-ligand affinities or factors that can explain why a specific protein surface region represents a viable binding site. While the interpretability of SBDD models aids in debugging and model enhancement, current efforts in this direction, such as those outlined in \cite{tubiana2022scannet, he2023nhgnn, rube2022prediction}, remain in their infancy and warrant further exploration.
\item \textbf{Biosecurity Concerns}: Structure-based drug design (SBDD) has revolutionized the development of targeted therapeutics by leveraging detailed structural information of biological molecules. However, this approach also introduces significant biosecurity concerns \cite{zhang2024foldmark, bloomfield2024ai, baker2024protein}. The accessibility of structural data, while beneficial for scientific collaboration, could be exploited to engineer harmful biological agents or enhance the virulence of existing pathogens. For instance, the Structural Genomics Consortium (SGC) \cite{williamson2000creating} openly shares three-dimensional structures of proteins, which, despite its intent to accelerate drug discovery, may inadvertently provide malicious actors with the information needed to design bioweapons. This dual-use dilemma underscores the necessity for stringent biosecurity measures and ethical guidelines to prevent the misuse of structural data in drug design.
\end{itemize}


\subsection{Opportunities}

\begin{itemize}
\item \textbf{Leverage Multimodal Datasets}: High-quality protein structure data remains limited; for example, CrossDocked and PDBBind datasets contain fewer than 10 thousand unique protein structures. In contrast, UniRef \cite{suzek2007uniref} boasts over 260 million protein sequences. As such, the incorporation of protein language models \cite{ferruz2022protgpt2, lin2022language, madani2020progen} trained on protein sequence data into structure-based drug design holds promise~\cite{ektefaie2023multimodal}. Additionally, textual data describing protein functions \cite{xu2023protst} and proteomics \cite{aslam2016proteomics} can be integrated into SBDD models.


\item \textbf{Incorporate Biological and Chemical Knowledge}: Integrating chemical and biomedical knowledge into model development has proven effective across various tasks. For instance, geometric symmetry is incorporated in equivariant neural networks, while molecular fragments are utilized to generate more realistic and valid molecules. Geometric deep learning stands to gain from the further infusion of domain knowledge.

\item \textbf{Build Comprehensive Benchmarks}: Standardized benchmarks offer dataset splits and evaluation tools, facilitating straightforward and robust comparison of SBDD models within a consistent framework~\cite{huang2021therapeutics,huang2022artificial}.

\item \textbf{Design Criteria Based on Clinical Endpoints}: Structure-based drug design is considered during the early stages of drug discovery and development. However, a palpable chasm exists between early-phase drug discovery and pre-clinical and clinical drug development~\cite{fu2022hint,fu2021probabilistic}. This can result in drug candidates faltering in clinical trials. Consequently, leveraging feedback from late drug development and using it to design novel design criteria to guide SBDD may increase therapeutic yield.

\item \textbf{Establish Foundation Models for SBDD}: Contemporary research on geometric deep learning methods for SBDD predominantly revolves around single-task models. However, with the emergence of general-purpose pre-trained models\cite{zhou2023comprehensive,huang2023zero,mcdermott2023structure,krishna2023generalized}, there is potential to develop unified foundation models that are compatible with a variety of data formats and tasks in SBDD.

\item \textbf{Consider a Broad Range of Design Tasks}: This survey examines geometric deep learning methods tailored for SBDD tasks, emphasizing small molecule drugs. Many methods are broadly applicable and can be adapted to other areas, such as antibody design \cite{kong2022conditional}, peptide design \cite{lin2024ppflow}, RNA design \cite{nori2024rnaflow}, and crystal material generation \cite{xie2021crystal}.
\end{itemize}

\section{Conclusion}
\label{sec:conclusion}
We systematically review geometric deep learning methods and applications for structure-based drug design. Our methodology involves categorizing existing research into five distinct categories based on the tasks they address. We present a comprehensive problem formulation for each task, summarizing noteworthy methods and delineating datasets and evaluation metrics. Considering both challenges and prospects for the field, we anticipate that this survey will facilitate a rapid comprehension of existing methodology and lay the groundwork for future structure-based drug design using geometric deep learning.



\bibliographystyle{IEEEtran}
\bibliography{main}






\begin{IEEEbiography}[{\includegraphics[width=1in,height=1.25in,clip,keepaspectratio]{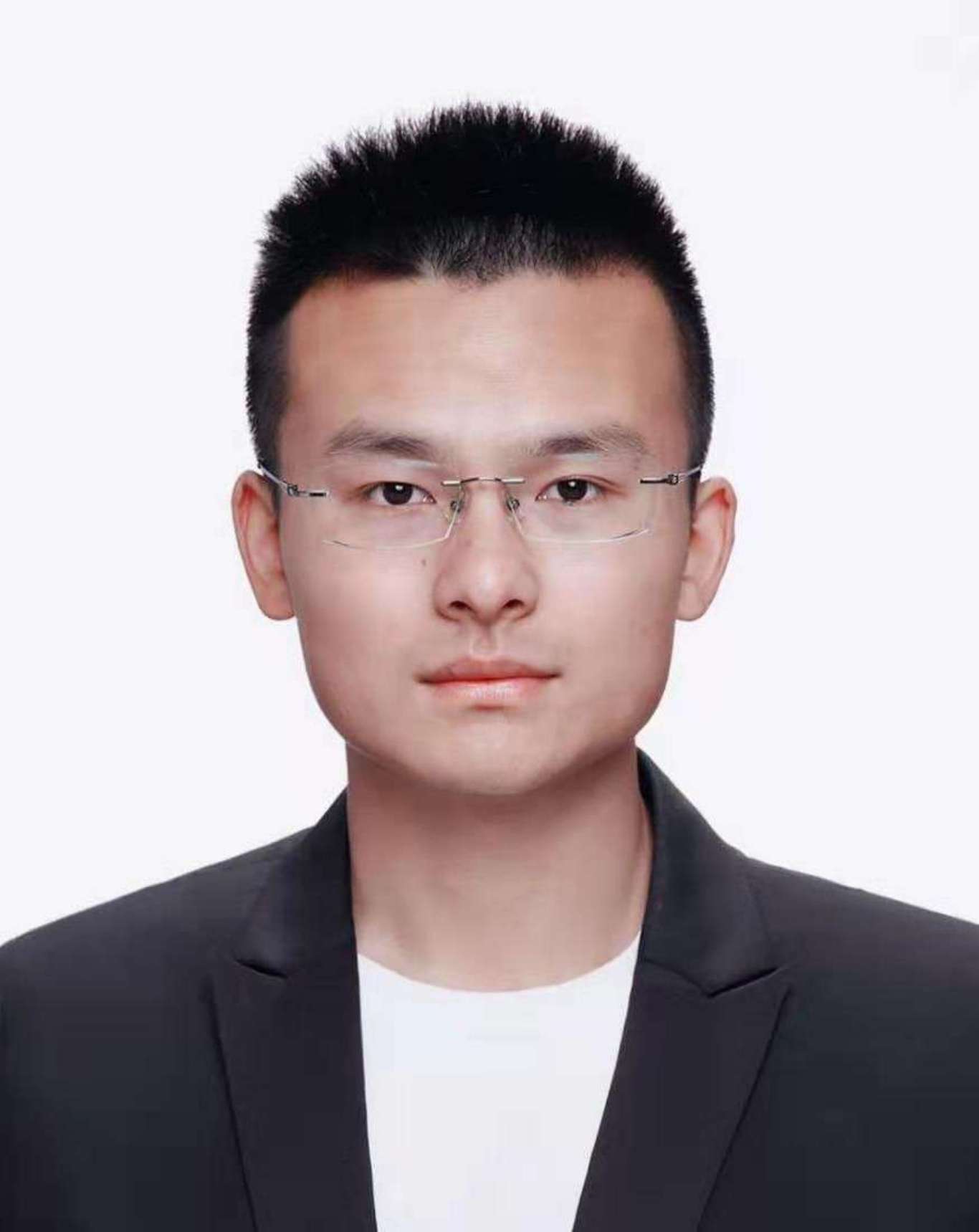}}] {Zaixi Zhang} received his B.S. degree from the Department of the Gifted Young, University of Science and Technology of China (USTC), in 2019, and his Ph.D. degree from the School of Computer Science and Technology at USTC in 2024. His research interests focus on graph representation learning and AI for drug discovery. He has published in top conferences and journals, including ICML, NeurIPS, CVPR, ICLR, TKDE, and TIFS. \end{IEEEbiography}
\vspace{-1em}
\begin{IEEEbiography}[{\includegraphics[width=1in,height=1.25in,clip,keepaspectratio]{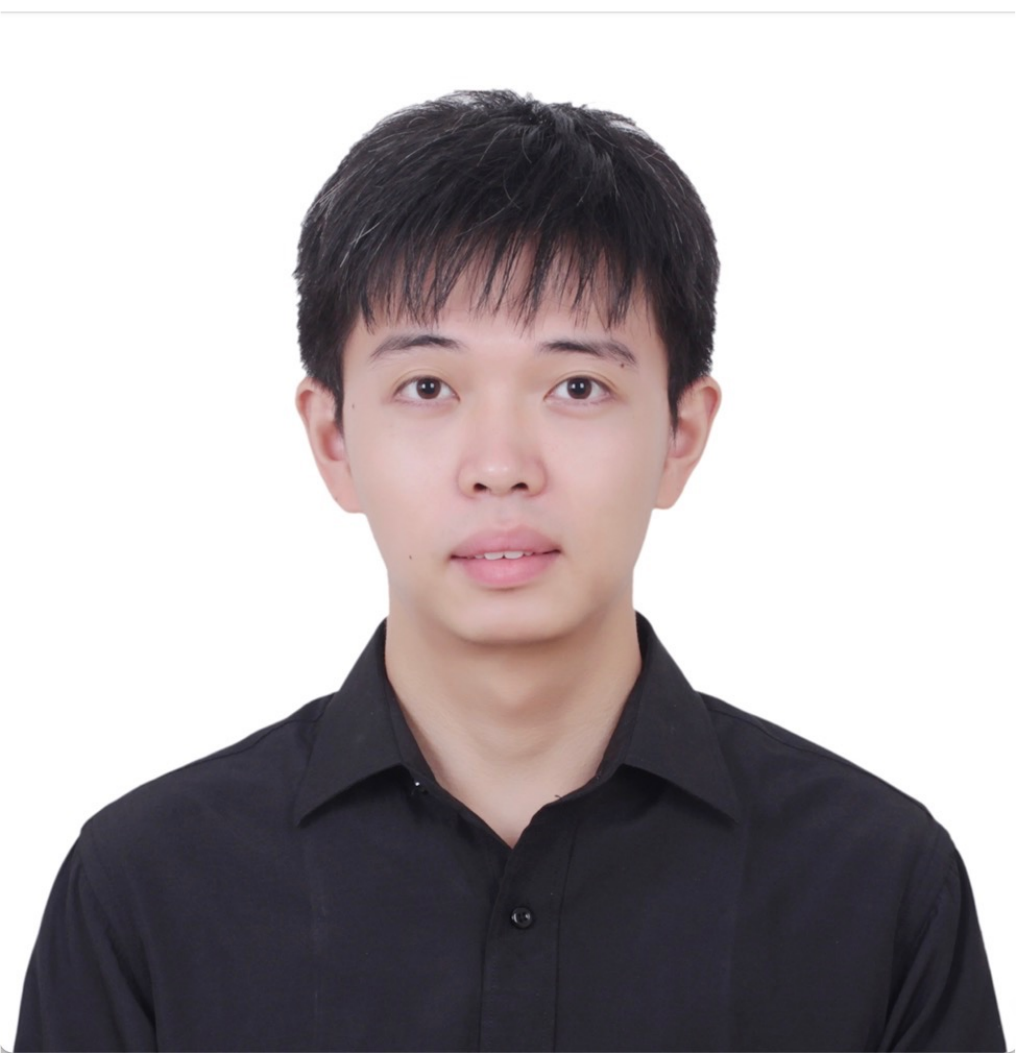}}]
        {Jiaxian Yan} is a doctoral candidate in the School of Data Science at the University of Science and Technology of China (USTC). His research focuses on large language models and AI-driven drug discovery. He has published in top conferences and journals, including NeurIPS and BiB.
\end{IEEEbiography}
\vspace{-1em}
\begin{IEEEbiography}[{\includegraphics[width=1in,height=1.25in,clip,keepaspectratio]{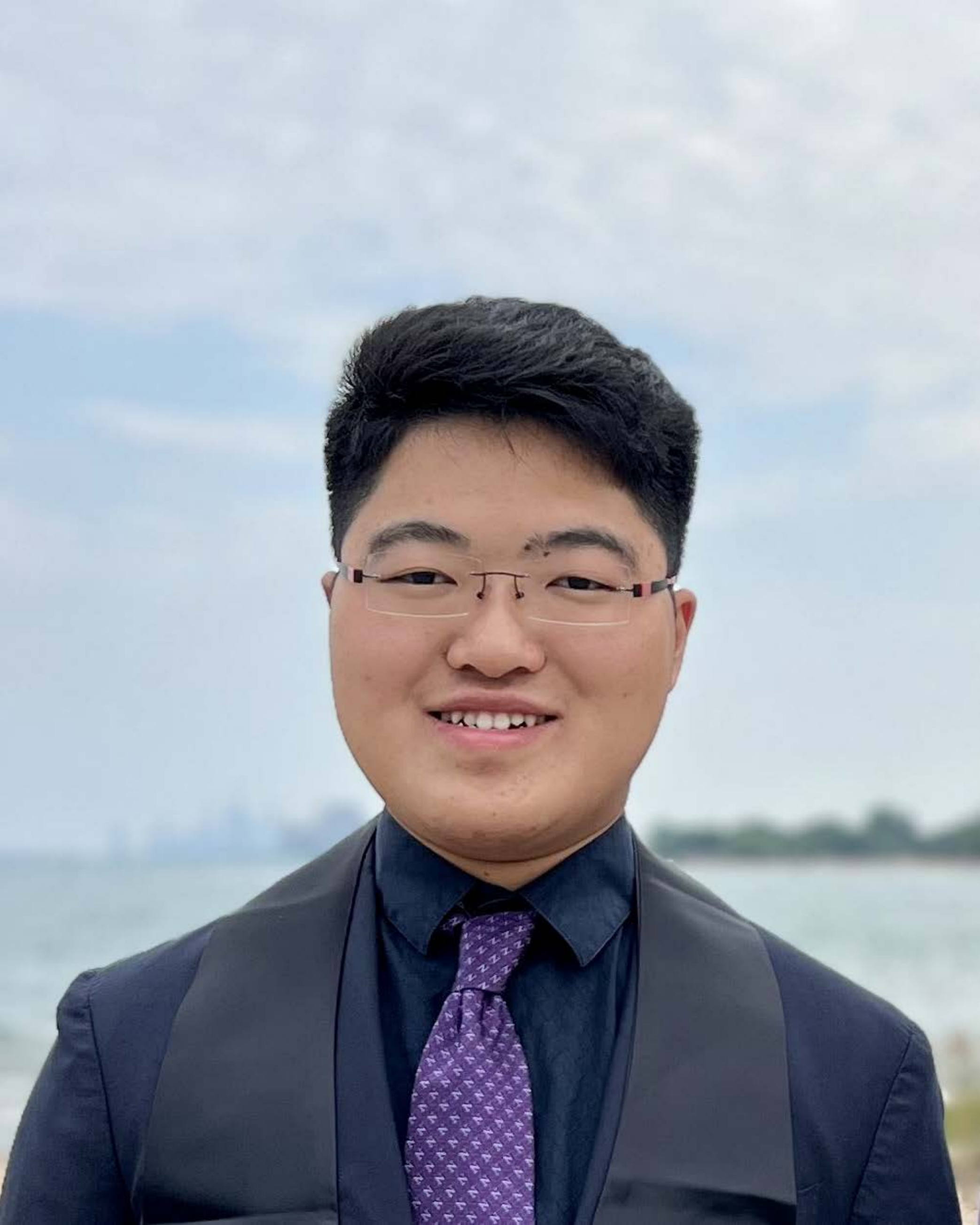}}]
        {Yining Huang} is a master’s student in biomedical informatics at Harvard University. His research focuses on AI for biology and drug design. 
\end{IEEEbiography}
\vspace{-1em}
\begin{IEEEbiography}[{\includegraphics[width=1in,height=1.25in,clip,keepaspectratio]{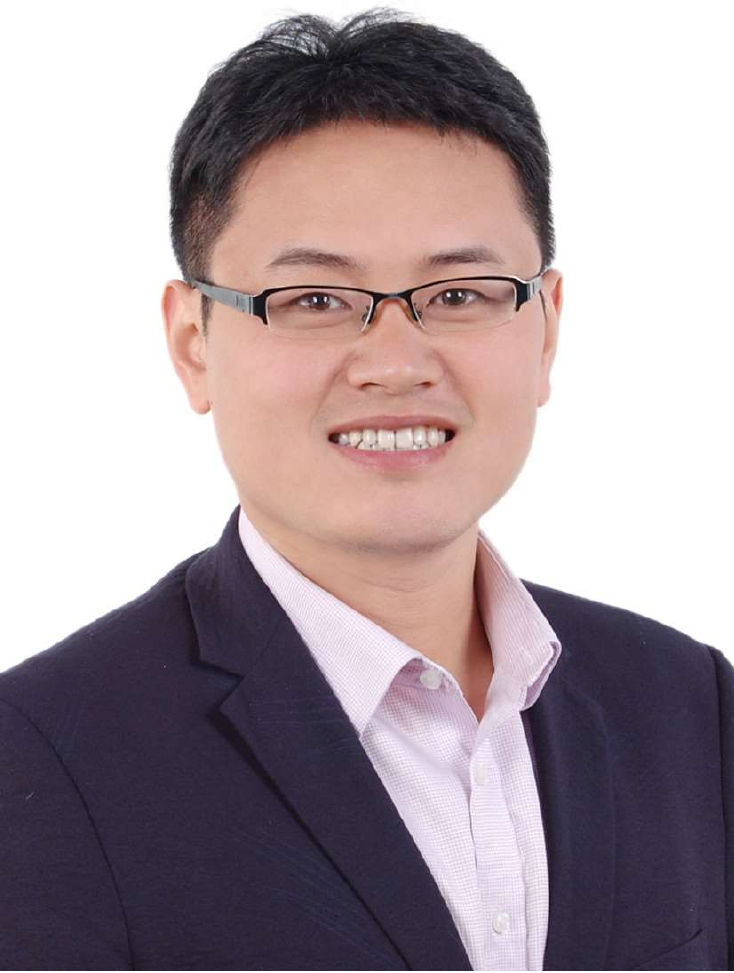}}]
	{Qi Liu} received the Ph.D. degree from University of Science and Technology of China (USTC), Hefei, China, in 2013. He is currently a Professor in the School of Computer Science and Technology at USTC. His general area of research is data mining and knowledge discovery. He has published prolifically in refereed journals and conference proceedings (e.g., TKDE, TOIS, KDD). He is an Associate Editor of IEEE TBD and Neurocomputing. He was the recipient of KDD' 18 Best Student Paper Award and ICDM'~11 Best Research Paper Award. He is a member of the Alibaba DAMO Academy Young Fellow. He was also the recipient of China Outstanding Youth Science Foundation in 2019. 
\end{IEEEbiography}
\vspace{-1em}
\begin{IEEEbiography}[{\includegraphics[width=1in,height=1.25in,clip,keepaspectratio]{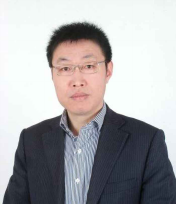}}]
	{Enhong Chen}(SM'07) is a professor and vice dean of the School of Computer Science at University of Science and Technology of China  (USTC) and an IEEE Fellow. He received the Ph.D. degree from USTC. His general area of research includes data mining and machine learning, social network analysis and recommender systems. He has published more than 100 papers in refereed conferences and journals, including IEEE Trans. KDE, IEEE Trans. MC, KDD, ICDM, NIPS, and CIKM. He was on program committees of numerous conferences including KDD, ICDM, SDM. He received the Best Application Paper Award on KDD-2008, the Best Student Paper Award on KDD-2018 (Research), the Best Research Paper Award on ICDM-2011 and Best of SDM-2015. His research is supported by the National Science Foundation for Distinguished Young Scholars of China.
\end{IEEEbiography}

\begin{IEEEbiography}[{\includegraphics[width=1in,height=1.25in,clip,keepaspectratio]{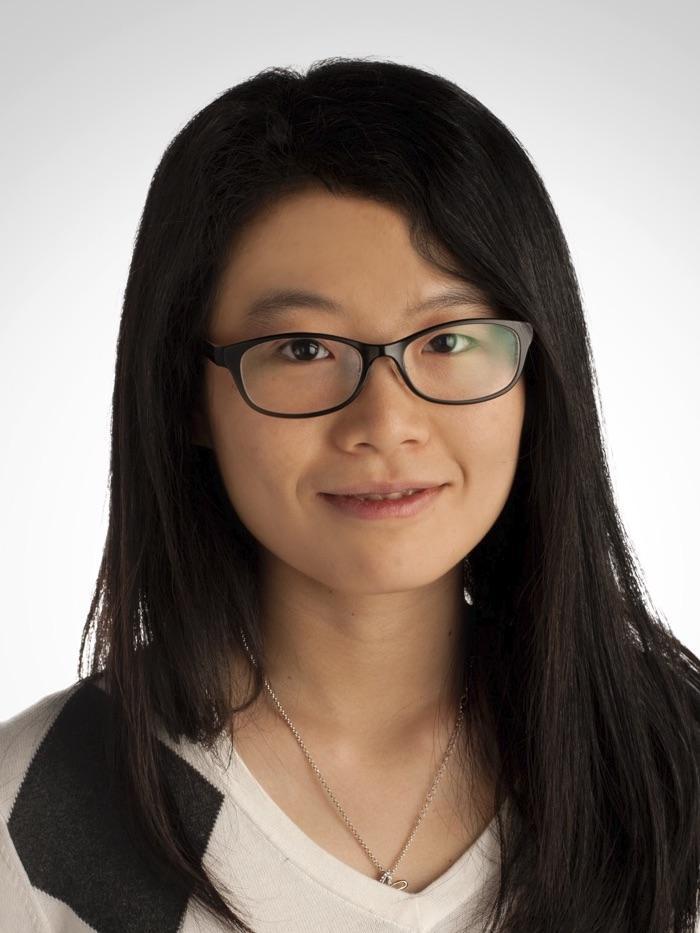}}]
	{Mengdi Wang} is an Associate Professor at Princeton University, holding joint appointments in the Department of Electrical and Computer Engineering and the Center for Statistics and Machine Learning. She also serves as Co-Director of the AI for Accelerating Invention initiative, which integrates artificial intelligence into engineering research to expedite technological advancements. Wang's research focuses on the theoretical foundations of machine learning and reinforcement learning, with applications spanning optimization, data science, and control systems. She earned her Ph.D. in Electrical Engineering and Computer Science from the Massachusetts Institute of Technology in 2013, under the supervision of Dimitri P. Bertsekas. Wang has received several accolades, including the Young Researcher Prize in Continuous Optimization from the Mathematical Optimization Society in 2016, the NSF CAREER Award in 2017, the Google Faculty Award in 2017, and recognition as one of MIT Technology Review's 35 Innovators Under 35 in China in 2018. She has served as a Program Chair for the International Conference on Learning Representations (ICLR) 2023 and as an associate editor for journals such as Operations Research and the Harvard Data Science Review. Wang's recent work includes developing language models to decode mRNA sequences, contributing to advancements in vaccine development and therapeutic applications.
\end{IEEEbiography} 

\begin{IEEEbiography}[{\includegraphics[width=1in,height=1.25in,clip,keepaspectratio]{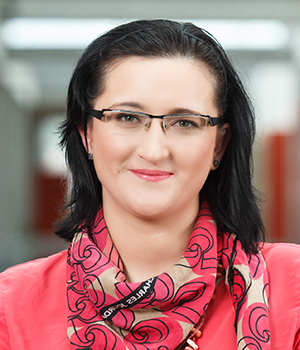}}]
	{Marinka Zitnik} is an Assistant Professor of Biomedical Informatics at Harvard University with additional appointments at the Kempner Institute for the Study of Natural and Artificial Intelligence, Broad Institute of MIT and Harvard, and Harvard Data Science. Zitnik investigates the foundations of AI to enhance scientific discovery and realize individualized diagnosis and treatment. The mission of her lab is to lay the foundations for AI to enhance the understanding of medicine, eventually enabling AI to learn and innovate on its own. Her research won several best paper and research awards, including the Kavli Fellowship of the National Academy of Sciences, awards from the International Society for Computational Biology, International Conference in Machine Learning, Bayer Early Excellence in Science, Amazon Faculty Research, Google Faculty Research, and Roche Alliance with Distinguished Scientists. Zitnik founded Therapeutics Data Commons, a global open-science initiative to access and evaluate AI across stages of development and therapeutic modalities, and she is also the faculty lead of the International AI4Science initiative.
\end{IEEEbiography}

\end{document}